\documentclass[preprint]{elsarticle}
\usepackage{graphicx} 
\usepackage{todonotes}
\usepackage{hyperref}
\usepackage{tikz}
\usepackage{csvsimple}
\usepackage{caption}
\usepackage{subcaption}
\usepackage[authormarkuptext=name,final]{changes}
\definechangesauthor[name={Reviewer \#1}, color=cyan]{1}
\definechangesauthor[name={Reviewer \#2}, color=red]{2}
\definechangesauthor[name={Author}, color=orange]{me}
\definechangesauthor[name={Structural}, color=green]{ed}
\setcommentmarkup{\todo[color=authorcolor]{#1}}
\usepackage{subcaption}
\usetikzlibrary{quantikz2}

\begin{document}

\begin{frontmatter}
    
\title{\replaced[id=me]{A Comparison of Quantum Compilers using a DAG-based or phase polynomial-based Intermediate Representation}{On the Architecture-Aware Synthesis of Intermediate Representations for Quantum Compilers}}
\author[1]{Arianne Meijer - van de Griend}
\ead{ariannemeijer@gmail.com}
\affiliation[1]{organization={Department of Computer Science, University of Helsinki},
addressline={Pietari Kalmin katu 5},
postcode={00560},
city={Helsinki},
country={Finland}}


\begin{abstract}
    In the NISQ era, where quantum computing is dominated by hybrid quantum algorithms, it is important for quantum circuits to be well-optimized to reduce noise from unnecessary gates. \deleted[id=me]{This paper describes current intermediate representations for architecture-aware quantum circuit synthesis.}
    We \replaced[id=me]{investigate}{compare} different \replaced[id=me]{phase polynomial-based compilation}{architecture-aware synthesis} strategies\added[id=me]{ to determine the current best practices and compare them} against the\added[id=me]{ DAG-based} Qiskit and TKET compilers. We \replaced[id=me]{find that phase polynomial-based compiling is very fast compared to DAG-based compiling. For long circuits, these compilers generate fewer CNOT gates than Qiskit or TKET, but for short circuits, they are}{show that the synthesis strategies do not find an optimal circuit, particularly for small circuits, where they seem to be} quite inefficient. We also show that supplementary algorithms \replaced[id=1]{such as}{like} Reverse Traversal and simulated annealing \added[id=me]{might }improve the generated CNOT count slightly, but \replaced[id=1]{the}{that} effect is \replaced[id=me]{negligable in most settings and generally not worth the additional compiler runtime. Instead, more sophisticated phase polynomial synthesis algorithms are needed}{much smaller than \added[id=1]{the improvements gained from using a better synthesis algorithm} \deleted[id=1]{the difference between improvements in synthesis algorithms}}.
    
\end{abstract}

\end{frontmatter}

\section{Introduction}
Small-scale quantum computing has recently become accessible to the public through cloud APIs \replaced[id=1]{such as}{like} AWS Braket~\cite{braket}, IBM Qiskit~\cite{Qiskit}, among others. With these new computing platforms comes the need for new compilation methods for quantum programs. Although some classical compilation methods can still be used for quantum computers, some compilation problems require a new solution due to the quantum nature of these devices. For example, most devices do not allow operation between arbitrary qubits. This is generally true of classical computers as well, but there we can solve the problem by copying bits. However, copying qubits is not possible due to the no-cloning theorem~\cite{wootters1982single}.

Similarly, optimizing quantum programs is a complex task. We somehow need to understand what the program is doing without implicitly calculating intermediate quantum states because simulating arbitrary quantum programs is infeasible on classical computers. It is difficult to gain these insights when studying a quantum circuit directly. In some cases, this approach can even hinder optimization~\cite{cowtan2020phase}. As such, it is crucial for the future of quantum computing to design appropriate intermediate representations that can be used for optimization in a way that is suitable for optimal machine code generation. 

In this paper, we 
\replaced[id=1]{
investigate the difference between compiling strategies that use a Directed Acyclic Graph (DAG) and those using phase polynomials as intermediate representation of the input quantum circuit. For the latter, we additionally investigate the effect of different aspects of the compilation process to determine a best-practice recommendation, namely the re-synthesis of trailing CNOTs, the use of Reverse Traversal to find a better qubit mapping, and the use of simulated annealing. These additional strategies have been investigated in limited settings before~\cite{meijer2022dynamic,gogioso2022annealing} but they have not yet been studied in a holistic setting.
}{give an overview of different intermediate representations that have been used in the literature and show experimentally that better architecture-aware synthesis algorithms are currently a key factor for the executability of medium-scale hybrid quantum algorithms.}

\added[id=2]{
The paper is structured as follows: we start with a short introduction to quantum compilers (\autoref{sec:compiler}) and an overview of intermediate representations from which a quantum circuit can be compiled (\autoref{sec:ir}). We then describe our research methods (\autoref{sec:methods}) and the results (\autoref{sec:results}). This is followed by a detailed discussion about the assumptions made during this research (\autoref{sec:discussion}). Lastly, we conclude from the obtained results and discuss future work (\autoref{sec:conclusion}).
}

\section{Compiling and transpiling for quantum computers}\label{sec:compiler}\deleted[id=ed, comment={new section}]{}
\added[id=1]{
By definition, a compiler is a computer program that transforms a program from one language to another. In the case of quantum computing, compilers typically transform quantum circuits into the native instruction set of the quantum computer. This instruction set does not need to be represented as a quantum circuit; it could be a pulse schedule for a superconducting quantum computer for example. If the compiler transforms the circuit to another circuit, this is called a transpiler. In that case, the native instruction set of the target quantum computer can be described in a quantum circuit, but there are restrictions as to which gates are allowed and how they can be used. In this paper, we will use these terms interchangeably. 
}

\begin{figure}
    \centering
    \includegraphics[width=0.5\linewidth]{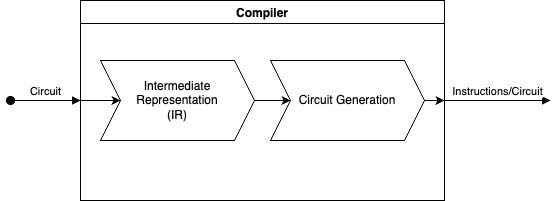}
    \caption{A simplified schematic representation of a quantum compiler.}
    \label{fig:comp}
\end{figure}

\added[id=1]{
Typically, a compiler transforms the input program using an intermediate representation that can accurately represent the function of the program in a manner that aids the transformation process without being restricted to the implementation details of the input program. For example, in classical computing, one can implement an iterative loop in a for-loop or a while-loop. An intermediate representation could be invariant under these two options. We will describe different intermediate representations for quantum circuits in \autoref{sec:ir}. A schematic overview of the workflow of a compiler is shown in \autoref{fig:comp}.
}

\added[id=1]{
A compiler needs to be efficient: it should find the shortest sequence of instructions as fast as possible. Particularly when computational resources are scarce, the quality of a compiler can determine whether a program is feasible to execute or not. This is also the case in quantum computing. One could argue that it is even more important for pre-fault-tolerant quantum computing than for classical computing because you cannot pause and continue a program as resources allow. Nor is it possible to use partially executed programs. Thus, a compiler needs to optimize the input program.
}

\added[id=1]{
Lastly, the compiler's output must conform to the native instruction set of the target quantum computer. The logical qubits in the circuit must be mapped to physical qubits on the hardware. Composite gates in the quantum circuit must be decomposed into gates directly supported by the target hardware. Additionally, if two-qubit gates are only supported between specific qubits, the program must be adjusted to only use the supported qubit pairs. The latter task is called the qubit routing problem, and it is known to be NP-hard~\cite{paler2021nisq}.
}

\section{Intermediate representations of quantum circuits}\label{sec:ir}
\replaced[id=ed]{
The capabilities of a compiler are largely influenced by the intermediate representation (IR) of the input circuit. Namely, if the IR cannot describe some property of the circuit, the compiler cannot leverage that property in optimization and circuit generation. In this section, we will describe two common IR structures: the Directed Acyclic Graph (DAG) and the phase polynomial. 
}{
In this section, we give a brief overview and introduction to architecture-aware synthesis for quantum algorithms from different intermediate representations. These intermediate representations follow a similar structure to the multi-qubit Pauli couplings (or PauliStrings) from which many hybrid quantum programs are constructed. Particularly circuit generated from a trotterized Hamiltonian, hardware-efficient ansatze, and QAOA algorithms follow this kind of structure.}


\subsection{Directed Acyclic Graph-based IR}\deleted[id=ed, comment={new section}]{}

\begin{figure}
    \centering
    \includegraphics[width=.5\linewidth]{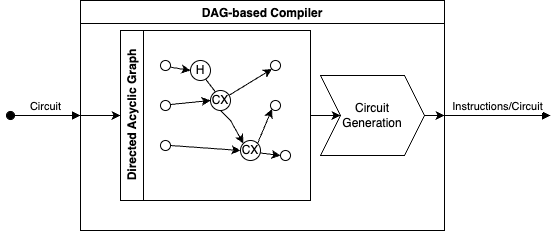}
    \caption{A simplified schematic representation of a DAG-based compiler.}
    \label{fig:dag}
\end{figure}

\replaced[id=1]{
As the name suggests, the DAG-based intermediate representation represents the circuit as a directed graph without loops. The graph describes the temporal dependencies of the gates in the circuit and which qubits they depend on (see \autoref{fig:dag}). Usually, compilers using a DAG-based IR (e.g. Qiskit~\cite{Qiskit} and TKET~\cite{tket}) are constructed from a series of \textit{passes} that change the DAG into an equivalent one. For example, a pass can remove instances where a gate is followed by its adjoint. 
}{
Aside from circuit optimization, the algorithms described in this section all try to solve the problem that not all qubits are connected with each other. Instead, the connections between qubits are described by a connectivity graph, and all multi-qubit gates in the circuit need to map to an edge in the connectivity graph. In this paper, we assume that the CNOT is the only multi-qubit gate in the circuit without loss of generality.
}

\replaced[id=1]{
Thus, this IR describes the chronological order of what the quantum computer needs to do and does not contain much information regarding the semantics of the circuit. This can make it hard to find global optimizations for the circuit. In particular, when adjusting a circuit to deal with the connectivity constraints of a target quantum computer, the DAG-based IR can only move qubits by inserting SWAP gates. 
}{
Traditional compilers, such as those in Qiskit~\cite{Qiskit} and TKET~\cite{tket}, take a given circuit and insert SWAP gates to move the qubits on the quantum computer, such that all CNOTs are allowed. This is done by representing the quantum circuit as a directed acyclic graph (DAG) that describes the causal dependencies of the gates in the circuit. 
However, this is a very difficult task~\cite{paler2021nisq} and this strategy can sometimes hinder optimization~\cite{cowtan2020phase}.
}

\deleted[id=1]{
Instead, the task of architecture-aware synthesis algorithms is to represent the circuit in an intermediate representation and recreate a completely new circuit from that representation such that all gates are immediately allowed by the connectivity graph and no SWAP gates are needed. 
}

\begin{figure}
    \centering
    \includegraphics[width=.9\linewidth]{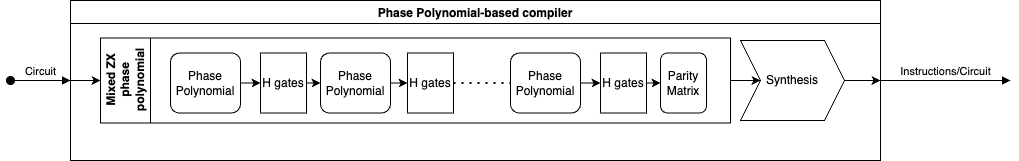}
    \caption{A simplified schematic representation of a phase polynomial-based compiler.}
    \label{fig:mixedrep}
\end{figure}

\subsection{Phase polynomial-based IR}\label{subsec:pp}
\added[id=ed, comment={Restructured the section}]{
Alternatively, one can try to represent the semantical meaning of the circuit. Ideally, one would describe the circuit with its unitary because that would not be scalable. Instead, we can rely on phase polynomials as a classically scalable representation. 
}

\added[id=1]{
In this paper, we use an intermediate representation that closely resembles a sequence of multi-qubit Z- and X-couplings called the mixed-ZX phase polynomial.
This is a composite representation that uses phase polynomials interleaved with a layer of Hadamard gates and ending with a parity matrix, as shown in \autoref{fig:mixedrep}. In the remainder of this section, we will briefly explain what these different components are, how they could make an intermediate representation on their own, and how they can be used to generate a new circuit. We will do so in a bottom-up manner, starting with the parity matrix. For further details, we refer the reader to the original papers.
}

\subsubsection{Parity matrix representation}
\replaced[id=ed]{
The parity matrix
}{One such intermediate representation is the parity matrix that} represents a sequence of CNOTs. Thus, we can cut the circuit on single qubit gates and recreate the CNOTs in between them. Many algorithms for architecture-aware synthesis of parity matrices exist~\cite{nash2020quantum,kissinger2020cnot,debrugiere2020quantum}, but here we will focus on the two algorithms that have been implemented for the results in this paper: RowCol~\cite{wu2019optimization} and PermRowCol~\cite{meijer2022dynamic}. All these algorithms use the fact that the effect of a sequence of CNOTs can be represented as a binary matrix, called the parity matrix~\cite{patel2008optimal}. In this matrix representation, the application of a CNOT gate corresponds to a row addition in the matrix. Thus, we can use constrained Gaussian Elimination to extract CNOTs from the parity matrix where we restrict the row operations to the edges in the coupling graph. 

However, in the constrained case, the strict column-wise ordering of Gaussian Elimination to go to the upper triangular form can be inefficient with respect to CNOT count. Thus, RowCol~\cite{wu2019optimization} does not eliminate the matrix to an upper triangular form but eliminates one row and corresponding column from the matrix at a time. Then, they make sure that the chosen row and column correspond to a qubit that does not cut the connectivity graph. That way, we know that the qubit is no longer needed and can be removed from the graph. 

Additionally, this strategy can be adjusted to pick a different qubit for the row and column, such that the parity matrix is reduced to a permutation matrix rather than an identity matrix. This corresponds to swapping the qubits on the quantum computer\added[id=me]{ which can be implemented as a classical post-processing step}. This is done using the PermRowCol algorithm~\cite{meijer2022dynamic} and it reduces the number of CNOT gates.

\subsubsection{Phase polynomial representation}
We can generalize the parity matrix representation to include single qubit Z rotations of arbitrary angles. This intermediate representation is called a phase polynomial~\cite{amy2014polynomialtime} and one can think of it as a sequence of multi-qubit Z coupling $e^{-i\alpha Z...Z}$, henceforth called phase gadgets.\added[id=me]{ For example, the phase polynomial of a CCZ gate is shown in \autoref{subfig:ir}.}
The qubits participating in these phase gadgets can be represented in a bitstring that behaves the same as the columns in the parity matrix representations. Thus, the trick is to generate CNOTs that are allowed by the architecture until a phase gadget only has a single qubit participating in it. Then, the phase gadget is a single qubit gate and it can be generated. The naive way of doing this is by synthesizing each phase gadget with its own ladder of CNOTs~\cite{cowtan2020phase} before the single qubit rotation and then undoing the ladder of CNOTs by recreating it in reverse\added[id=1]{ as shown in \autoref{subfig:naive} for the CCZ gate}. But it is \replaced[id=me]{clearly}{of course} highly inefficient, and we want subsequence CNOT ladders to cancel each other out.

\begin{figure}
    \centering
    \begin{subfigure}[b]{\linewidth}
    \centering
    $$CCZ = e^{-i\frac{\pi}{4}\left(ZII + IZI + IIZ - ZIZ - IZZ - ZZI + ZZZ\right)}$$
\caption{Phase polynomial of a CCZ gate.}\label{subfig:ir}
    \end{subfigure}
    \hfill
    \begin{subfigure}[b]{\linewidth}
    \centering
    \begin{tikzpicture}
    \node[scale=.5] {
    \begin{quantikz}
&\gate{T}\gategroup[1,steps=1,style={dashed,rounded corners,fill=blue!20, inner xsep=0pt, inner ysep=0pt},background,label style={label position=below, anchor=north,yshift=-0.2cm}]{$e^{-i\frac{\pi}{4}ZII} $}&\targ{}\gategroup[3,steps=7,style={dashed,rounded corners,fill=blue!20, inner xsep=0pt, inner ysep=0pt},background,label style={label position=below, anchor=north,yshift=-0.2cm}]{$e^{i\frac{\pi}{4}ZIZ} $}&\ctrl{1}&&&&\ctrl{1}&\targ{}&&&&\ctrl{1}\gategroup[2,steps=3,style={dashed,rounded corners,fill=blue!20, inner xsep=0pt, inner ysep=0pt},background,label style={label position=below, anchor=north,yshift=-0.2cm}]{$e^{i\frac{\pi}{4}ZZI} $}&&\ctrl{1}&\ctrl{1}\gategroup[3,steps=5,style={dashed,rounded corners,fill=blue!20, inner xsep=0pt, inner ysep=0pt},background,label style={label position=below, anchor=north,yshift=-0.2cm}]{$e^{-i\frac{\pi}{4}ZZZ} $}&&&&\ctrl{1}&\\[10pt]
&\gate{T}\gategroup[1,steps=1,style={dashed,rounded corners,fill=blue!20, inner xsep=0pt, inner ysep=0pt},background,label style={label position=below,anchor=north,yshift=-0.2cm}]{$e^{-i\frac{\pi}{4}IZI} $}&\ctrl{-1}&\targ{}&\ctrl{1}&&\ctrl{1}&\targ{}&\ctrl{-1}&\ctrl{1}\gategroup[2,steps=3,style={dashed,rounded corners,fill=blue!20, inner xsep=0pt, inner ysep=0pt},background,label style={label position=below, anchor=north,yshift=-0.2cm}]{$e^{i\frac{\pi}{4}IZZ} $}&&\ctrl{1}&\targ{}&\gate{T^\dagger}&\targ{}&\targ{}&\ctrl{1}&&\ctrl{1}&\targ{}& \\[10pt]
&\gate{T}\gategroup[1,steps=1,style={dashed,rounded corners,fill=blue!20, inner xsep=0pt, inner ysep=0pt},background,label style={label position=below,anchor=north,yshift=-0.2cm}]{$e^{-i\frac{\pi}{4}IIZ} $}&&&\targ{}&\gate{T^\dagger}&\targ{}&&&\targ{}&\gate{T^\dagger}&\targ{}&&&&&\targ{}&\gate{T}&\targ{}&&
\end{quantikz}
};
\end{tikzpicture}
\caption{Naive decomposition of the CCZ gate.}\label{subfig:naive}
    \end{subfigure}
    \begin{subfigure}[b]{\linewidth}
    \centering
    \begin{tikzpicture}
    \node(box1)[scale=.5] {
    \begin{quantikz}
        &\gate[3]{e^{-i\frac{\pi}{4}\left(ZII + IZI + IIZ - ZIZ - IZZ - ZZI + ZZZ\right)}}&\\
        &&\\
        &&
    \end{quantikz}
    =
    \begin{quantikz}
&\gate{T}&&&\gate[3]{e^{-i\frac{\pi}{4}\left(- ZIZ - IZZ - ZZI + ZZZ\right)}}&\\
&\gate{T}&\targ{}&\targ{}&& \\
&\gate{T}&\ctrl{-1}&\ctrl{-1}&&
    \end{quantikz}};
    \node(box2)[scale=.5, yshift=-80pt] {
    = 
    \begin{quantikz}
&\gate{T}&&\gate[3]{e^{-i\frac{\pi}{4}\left(- ZIZ - IZI - ZZZ + ZZI\right)}}&&\\
&\gate{T}&\targ{}&&\targ{}& \\
&\gate{T}&\ctrl{-1}&&\ctrl{-1}&
\end{quantikz}
    = 
    \begin{quantikz}
&\gate{T}&&&\gate[3]{e^{-i\frac{\pi}{4}\left(- ZIZ - ZZZ + ZZI\right)}}&&\\
&\gate{T}&\targ{}&\gate{T^\dagger}&&\targ{}& \\
&\gate{T}&\ctrl{-1}&&&\ctrl{-1}&
\end{quantikz}
    };
    \end{tikzpicture}
        \caption{CNOT generation during phase polynomial synthesis.}\label{subfig:cnot}
    \end{subfigure}
    \begin{subfigure}[b]{\linewidth}
    \centering
    \begin{tikzpicture}
    \node(box1)[scale=.5] {
    \begin{quantikz}
&\gate{T}\gategroup[1,steps=1,style={dashed,rounded corners,fill=blue!20, inner xsep=0pt, inner ysep=0pt},background,label style={label position=below,anchor=north,yshift=-0.2cm}]{ZII}&&&\ctrl{1}&&&&\targ{}&\ctrl{1}&&&\gategroup[3,steps=6,style={dashed,rounded corners,fill=blue!20, inner xsep=0pt, inner ysep=0pt},background,label style={label position=below, anchor=north,yshift=-0.2cm}]{\text{Trailing CNOTs}}&\ctrl{1}&\targ{}&&\ctrl{1}&&\\[10pt]
&\gate{T}\gategroup[1,steps=1,style={dashed,rounded corners,fill=blue!20, inner xsep=0pt, inner ysep=0pt},background,label style={label position=below,anchor=north,yshift=-0.2cm}]{IZI}&\targ{}&\gate{T^\dagger}\gategroup[1,steps=1,style={dashed,rounded corners,fill=blue!20, inner xsep=0pt, inner ysep=0pt},background,label style={label position=below,anchor=north,yshift=-0.2cm}]{IZZ}&\targ{}&\gate{T}\gategroup[1,steps=1,style={dashed,rounded corners,fill=blue!20, inner xsep=0pt, inner ysep=0pt},background,label style={label position=below,anchor=north,yshift=-0.2cm}]{ZZZ}&\targ{}&\gate{T^\dagger}\gategroup[1,steps=1,style={dashed,rounded corners,fill=blue!20, inner xsep=0pt, inner ysep=0pt},background,label style={label position=below,anchor=north,yshift=-0.2cm}]{ZZI}&\ctrl{-1}&\targ{}&\ctrl{1}&&\ctrl{1}&\targ{}&\ctrl{-1}&\targ{}&\targ{}&\targ{}&\\[10pt]
&\gate{T}\gategroup[1,steps=1,style={dashed,rounded corners,fill=blue!20, inner xsep=0pt, inner ysep=0pt},background,label style={label position=below,anchor=north,yshift=-0.2cm}]{IIZ}&\ctrl{-1}&&&&\ctrl{-1}&&&&\targ{}&\gate{T^\dagger}\gategroup[1,steps=1,style={dashed,rounded corners,fill=blue!20, inner xsep=0pt, inner ysep=0pt},background,label style={label position=below,anchor=north,yshift=-0.2cm}]{ZIZ}&\targ{}&&&\ctrl{-1}&&\ctrl{-1}&
    \end{quantikz}
    };
    \end{tikzpicture}
    \caption{Synthesized CCZ from the phase polynomial}\label{subfig:full}
    \end{subfigure}
    \begin{subfigure}[b]{\linewidth}
    \centering
    \begin{tikzpicture}
    \node(box1)[scale=.5] {
    \begin{quantikz}
\lstick{\ket{q_1}}&\gate{T}&&&\ctrl{1}&&&&\targ{}&\ctrl{1}&&&&\rstick{\ket{q_2}}\\
\lstick{\ket{q_2}}&\gate{T}&\targ{}&\gate{T^\dagger}&\targ{}&\gate{T}&\targ{}&\gate{T^\dagger}&\ctrl{-1}&\targ{}&\ctrl{1}&&\ctrl{1}&\rstick{\ket{q_1}}\\
\lstick{\ket{q_3}}&\gate{T}&\ctrl{-1}&&&&\ctrl{-1}&&&&\targ{}&\gate{T^\dagger}&\targ{}&\rstick{\ket{q_3}}
    \end{quantikz}
    };
    \end{tikzpicture}
    \caption{Synthesized CCZ from \autoref{subfig:full} with trailing CNOTs re-synthesized up-to-permutation.}\label{subfig:prc}
    \end{subfigure}
    \caption{Example architecture-aware phase polynomial synthesis of a CCZ gate restricted to a 3-qubit line topology. The phase polynomial representation is shown in \autoref{subfig:ir}. The resulting circuit from using naive decomposition is shown in \autoref{subfig:naive} and using a more sophisticated approach in \autoref{subfig:full}. The generation of gates during synthesis and how it updates the phase polynomial is shown in \autoref{subfig:ir} and the CCZ with trailing CNOTs re-synthesized while reallocating the input qubits is shown in \autoref{subfig:prc}.}
    \label{fig:synth-example}
\end{figure}

The phase polynomial synthesis algorithms discussed in this paper do this by assuming that every CNOT that is generated is canceled out and each multi-qubit Z coupling is updated accordingly. The CNOTs after the last phase gadget \deleted[id=me]{cannot be canceled out, so they} are aggregated at the ends as trailing CNOTs. \replaced[id=1]{
The CNOTs are generated as needed, commuted through the phase gadgets, and aggregated at the end as shown in \autoref{subfig:cnot}. As a result, the trailing CNOTs are unoptimized (see \autoref{subfig:full}) and might even contain sequences of CNOTs acting on the same two qubits. This happens when a CNOT is assumed to cancel out, but was instead needed for the next phase gadget.
}{At some point, it is possible that any of these CNOTs were actually required later on and should not have been canceled out. In that case, it will be generated again and its counterpart will be aggregated in the trailing CNOTs.} This is not a problem because we can represent these trailing CNOTs as a parity matrix and optimize them by resynthesizing them\added[id=1]{ as shown in \autoref{subfig:prc}}.

So how do we pick which CNOTs to generate? One way to do this is to generate the parities in order of Gray-code~\cite{amy2018controlled} such that every subsequent parity requires a minimum number of CNOTs. However, this ordering does not take into account the connectivity graph. Although this can be overcome by adding extra CNOTs in between~\cite{nash2020quantum}, it is better to follow the Gray-code structure by restricting the ordering to follow the ``ends`` of the connectivity graph (non-cutting vertices)~\cite{meijer-vandegriend2020architectureaware}. However, this is still assuming that every bitflip in a parity has the same CNOT cost, but this is not true. Thus, \cite{vandaele2022phase} proposed a look-ahead scheme for choosing the order in which the phase gadgets (multi-qubit Z couplings) should be synthesized. In the remainder of this paper, we will refer to the method from \citet{meijer-vandegriend2020architectureaware} as Steiner-GraySynth (\texttt{sgs}) and the one from \citet{vandaele2022phase} as ParitySynth (\texttt{ps}).

\subsubsection{Mixed ZX-phase polynomial representation}
The previous intermediate representations are not universal for quantum computing. In fact, they are classically simulable~\cite{amy2014polynomialtime}\added[id=me, comment={extra ref}]{}. However, we use these principles in the intermediate representation called mixed ZX phase polynomials~\cite{gogioso2022annealing}. 
As the name suggests, this intermediate representation uses the concept of X phase polynomials. These are phase polynomials in the Hadamard basis, rather than the computational basis. Thus, it is a sequence of multi-qubit X couplings $e^{-i \alpha X...X}$, henceforth called X phase gadgets. Thus, we represent the circuit as CNOTs, $R_Z(\alpha)$, and $R_X(\alpha)$, we write the single qubit gates as Z- and X-phase polynomials and then move the CNOTs to the end of the circuit and aggregate them there. Then, we can synthesize subsequent phase polynomials using the phase polynomial synthesis algorithms described before and then synthesize the trailing CNOTs using the parity matrix synthesis algorithms.

\subsubsection{Supplemental algorithms}

\begin{figure}
    \centering
    \includegraphics[width=0.6\linewidth]{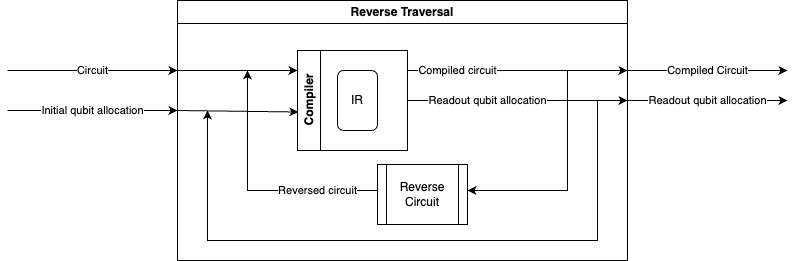}
    \caption{A schematic representation of the reverse traversal strategy.}
    \label{fig:rt}
\end{figure}
\begin{figure}
    \centering
    \includegraphics[width=0.5\linewidth]{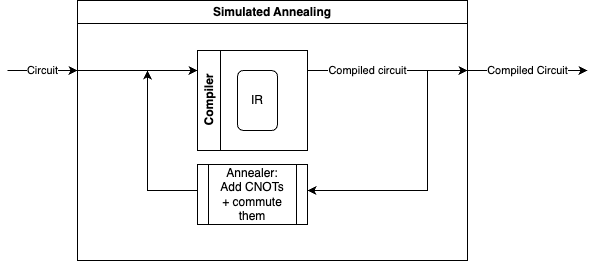}
    \caption{A schematic representation of using simulated annealing in the compiler.}
    \label{fig:sa}
\end{figure}

Additionally, these algorithms can be supplemented with heuristical algorithms that attempt to improve the CNOT count. If the algorithm reallocates the qubits during synthesis, as in the case of \replaced[id=1]{
PermRowCol~\cite{meijer2022dynamic} }{PermRowCol Meijer - van de Griend and Li ~\cite{meijer2022dynamic} }, 
we can use the Reverse Traversal Strategy from \replaced[id=1]{Qiskit's}{Qiskti} SABRE~\cite{li2019tackling} to try to find an optimal qubit mapping by recompiling the circuit and its reverse\added[id=1]{ as illustrated in \autoref{fig:rt}}.

One could also try to find common parities in the phase polynomials using simulated annealing~\cite{gogioso2022annealing}. This strategy adds CNOTs to the start of the circuit such that the synthesis algorithm generates fewer CNOTs\added[id=1]{, as shown in \autoref{fig:sa}}. Intuitively, the annealer transforms the input space of the quantum circuit to one that is easier to synthesize.

\section{Methods}\label{sec:methods}
\added[id=1]{
In this paper, we want to investigate how DAG-based compiling compares to phase polynomial-based compiling. However, both of these strategies have many different sub-procedures that can be adjusted.
DAG-based methods can use a very different sequence of compiler passes depending on the compiler, but they are only comparatively tested as a whole~\cite{tket,li2019tackling}. 
Conversely, as described in \autoref{subsec:pp}, phase polynomial-based methods have a well-defined sequence of steps, but each step is investigated on its own without investigating the big picture. In the case of the former, the best sequence of compiler passes is often well-documented, but for phase polynomial-based compiling, no such best-practices are defined.
}

\added[id=1]{
Thus, we first want to determine what the best strategy for phase polynomial based compiling is. Given a method for phase polynomial synthesis, we ask the following questions:
}
\begin{enumerate}
    \item[RQ1] \added[id=1]{Does the re-synthesis of trailing CNOTs improve performance?}
    \item[RQ2] \added[id=1]{Does Reverse Traversal improve performance?}
    \item[RQ3] \added[id=1]{Does simulated annealing improve performance?}
    \item[RQ4] \added[id=1]{Does a combination of Reverse Traversal and simulated annealing improve performance? }
\end{enumerate}

\added[id=1]{
Once these questions have been answered, we decide on a best-practice to use in the comparison against the DAG-based compiling methods. Such that we can answer our main research question:
}
\begin{enumerate}
    \item[RQ5] \added[id=1]{What is the performance difference between DAG-based compiling and phase polynomial-based compiling?}
\end{enumerate}
\added[id=1]{
Here, we investigate the performance in terms of CNOT count and compiler runtime. A good compiler creates small circuits in short amount of time. 
}

\begin{figure}
    \centering
    \includegraphics[width=0.5\linewidth]{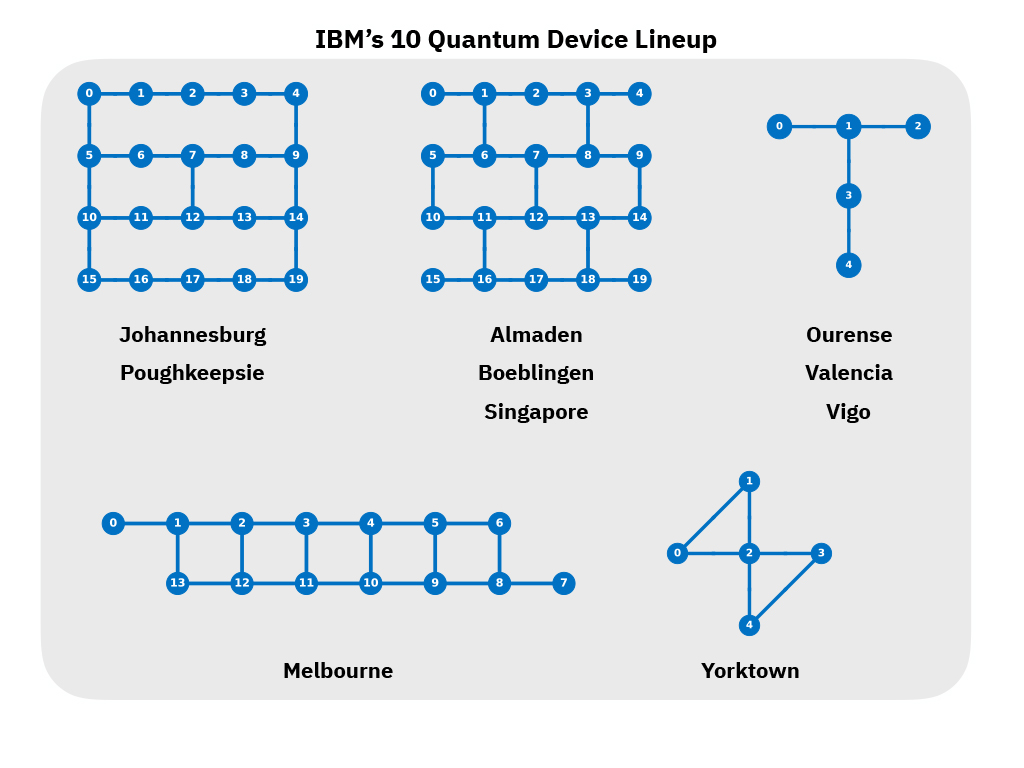}
    \caption{Connectivity graphs for the devices used in the experiments}
    \label{fig:archs}
\end{figure}

To give an indication of the effect of these different \replaced[id=me]{strategies}{algorithms} on the compiled machine code for different quantum computer architectures, we have run various experiments. We used five different publicly available IBM architectures: $20$-qubit Johannesburg, $20$-qubit Singapore, $14$-qubit Melbourne, $5$-qubit Yorktown, and $5$-qubit Valencia. \added[id=me]{The different connectivity graphs are shown in \autoref{fig:archs}.}

For each of these devices, we generated $80$ mixed ZX-phase polynomials with $1$, $2$, $5$, $10$, $50$, and $100$ random phase gadgets. To force that these phase gadgets represent complicated interactions between the qubits, we restricted the number of qubits in each phase gadget interaction to be at least $\sqrt{n}$, where $n$ is the number of available qubits. 
\added[id=me]{Such that this synthetic dataset resembles common NISQ algorithms at the edge of and just beyond what is executable on quantum computers at the time of writing.}
Then, we naively generated a circuit from those phase gadgets, synthesizing the phase gadgets one by one following \citet{cowtan2020phase} while assuming full connectivity between the qubits. This naively generated circuit served as the input for the different compiling methods.

As \replaced[id=me]{DAG-based compilers}{non-synthesis baselines}, we have used the Qiskit transpiler and the TKET compiler, as these are two popular compilers used in practice when researchers want to execute their quantum circuit on a device. \added[id=1]{
To give a fair comparison, we use the settings that are documented to give the best performance. This means that
} \replaced[id=1]{for}{For} Qiskit, we used Qiskit's \texttt{transpile()}`` method with optimization level 3\footnote{\url{https://docs.quantum.ibm.com/api/qiskit/0.39/qiskit.compiler.transpile\#qiskitcompilertranspile}}. \replaced[id=1]{And for}{For} TKET, we used the default compilation pass with optimization level 2, recreated from documentation\footnote{\url{https://cqcl.github.io/pytket-qiskit/api/index.html\#default-compilation}} for the target architectures \added[id=me]{as the target architectures were nor directly supported by TKET}. \added[id=1]{Note that even though both these settings are called \textit{"optimization level"}, they do not refer to the same optimization strategy nor do they follow the same scale.}

For the phase polynomial\replaced[id=me]{-based compiling methods}{ synthesis}, we have used the naive synthesis strategy \added[id=me]{from~\citet{cowtan2020phase} }(\texttt{naive}), the algorithm from \citet{vandaele2022phase} (\texttt{ps}), and the algorithm from \citet{meijer-vandegriend2020architectureaware} (\texttt{sgs}). The CNOT circuit aggregated at end of the phase polynomial can be left as is\deleted[id=me]{ (\texttt{naive}}, or re-synthesized using RowCol~\cite{wu2019optimization} (\texttt{rc}) or PermRowCol~\cite{meijer2022dynamic} (\texttt{prc}). The latter \replaced[id=me]{can also reallocate}{also reallocates} the qubits on the device\added[id=me]{.} 

\begin{figure}
    \centering
    \begin{subfigure}[b]{0.4\textwidth}
    \centering
    \includegraphics[width=1\linewidth]{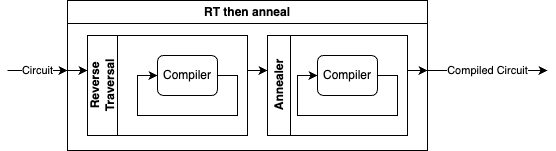}
    \caption{\texttt{RT then anneal}}
    \end{subfigure}
    \hfill
    \begin{subfigure}[b]{0.25\textwidth}
    \centering
    \includegraphics[width=1\linewidth]{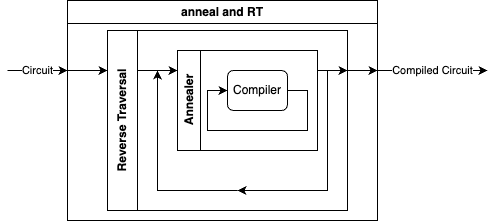}
    \caption{\texttt{anneal and RT}}
    \end{subfigure}
    \hfill
    \begin{subfigure}[b]{0.24\textwidth}
    \centering
    \includegraphics[width=1\linewidth]{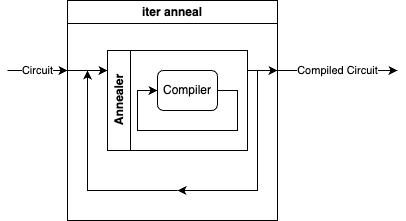}
    \caption{\texttt{iter anneal}}
    \end{subfigure} 
    \caption{Schematic representation of the two methods to combine Reverse Traversal with simulated annealing and the baseline of iteratively rerunning the annealer.}
    \label{fig:combined-fig}
\end{figure}

Then, the synthesis algorithms could still be improved with the supplemental algorithms. For this, we have used the Reverse Traversal Strategy~\cite{li2019tackling} (\texttt{RT}) that can only be used in combination with PermRowCol~\cite{meijer2022dynamic} and attempts to find a better initial qubit mapping. We also used the simulated annealing strategy from \citet{gogioso2022annealing} (\texttt{anneal}). Additionally, we can combine these supplemental algorithms in different ways. \replaced[id=me]{We}{First, we} can run the annealer after the Reverse Traversal has found a good initial mapping (\texttt{RT then anneal}), or we can use the Reverse Traversal in every step of the simulated annealer (\texttt{anneal and RT}). To see if the difference in performance might be due to the restarting of the simulated annealer, we have also \replaced[id=me]{included a strategy}{done an experiment} where we iteratively restart the simulated annealer multiple times and take the best solution (\texttt{iter anneal}). \added[id=me]{Note that we include the CNOTs added by the annealer in the trailing CNOTs to be re-synthesized. The difference between these combined algorithms is shown in \autoref{fig:combined-fig}.}

We \replaced[id=me]{used}{ran the experiments that use the supplemental algorithms using} $10$ iterations for Reverse Traversal and \replaced[id=me]{i}{I}terative \replaced[id=me]{a}{A}nnealing. Following \citet{gogioso2022annealing}, we used $5$ CNOT blocks and a linear schedule for simulated annealing.
Additionally, we used only $1000$ iterations for the simulated annealing. We know from \citet{gogioso2022annealing} that this is the lower bound of iterations from which the simulated annealing started to converge. \replaced[id=me]{I}{We will show and discuss later that i}ncreasing the number of iterations to e.g. $5000$ would not have been possible due to the additional compilation time.

\added[id=me]{
The full code base, including raw results and instructions for replication, can be found on Github
}\footnote{\url{https://github.com/Aerylia/pauliopt/tree/paper/reverse-traversal}}.

\section{Results}\label{sec:results}
\added[id=ed,comment={Rewrote results section completely}]{
In the previous section, we defined $5$ research questions. In this section, we will investigate each of these questions in their own subsection.
}

\subsection{The effect of CNOT re-synthesis}\label{subsec:cnot}
\added[id=ed]{
First, we investigate the effect of CNOT re-synthesis by plotting the CNOT count and compiler runtime against the number of phase gadgets for each phase polynomial synthesis method with different strategies for trailing CNOT optimization.
}

\added[id=ed]{
Note that for naive decomposition, the re-synthesis of trailing CNOTs will have at best re-synthesize the same CNOTs due to how the naive decomposition generates the CNOTs. Thus, the analysis for the naive decomposition is omitted, but the naive decomposition is shown as a baseline for the other phase polynomial synthesis methods. These alternative methods should result in smaller circuits.
}

\begin{figure}
    \centering
    \includegraphics[width=0.75\linewidth]{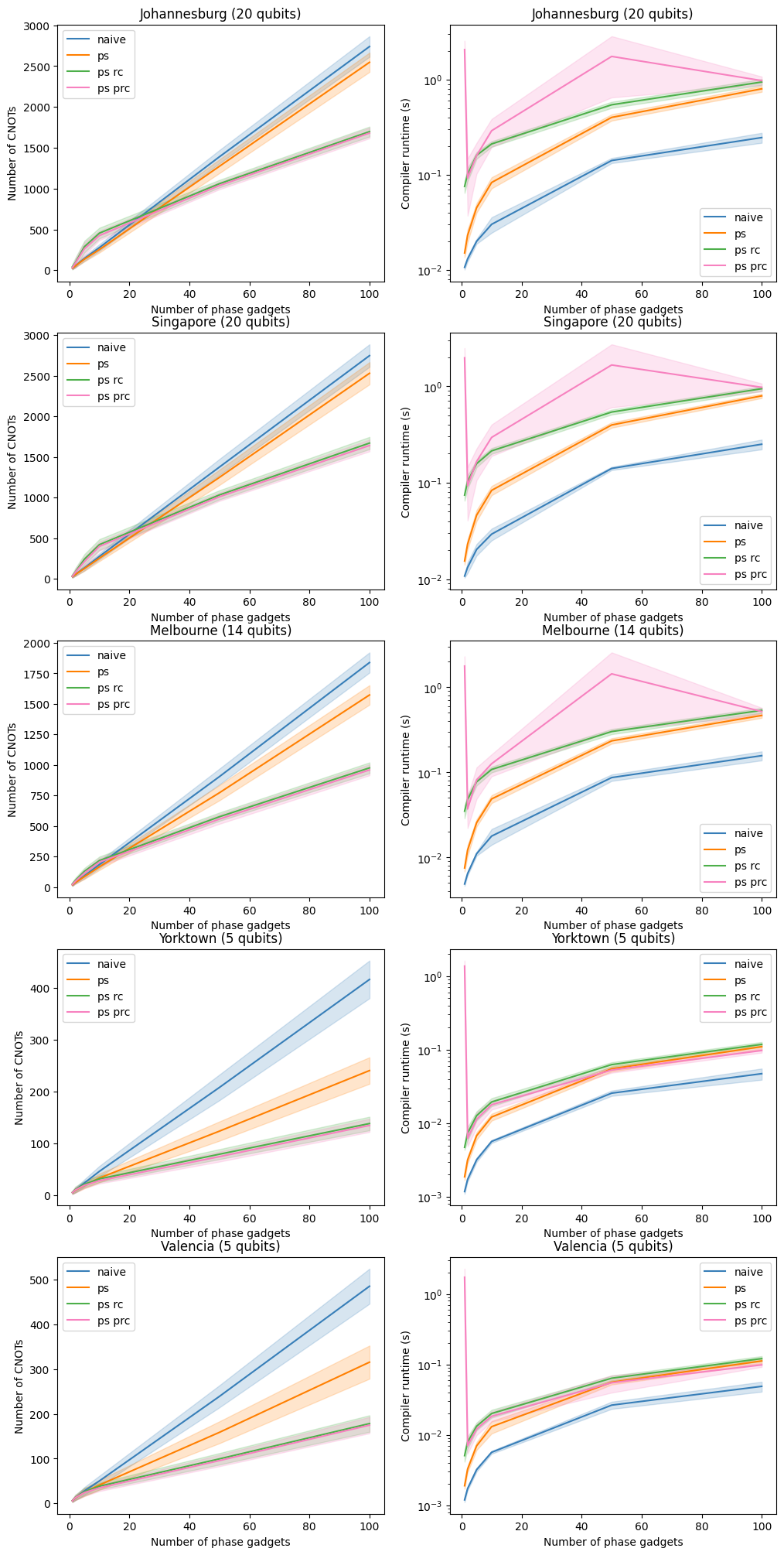}
    \caption{A comparison of CNOT count (left) and compiler runtime (right) for circuits compiled using ParitySynth~\cite{vandaele2022phase} without re-synthesizing trailing CNOTs (\textit{ps}), with trailing CNOTs re-synthesized with RowCol~\cite{wu2019optimization} (\texttt{ps rc}) or PermRowCol~\cite{meijer2022dynamic} (\texttt{ps prc}), or compiled using naive decomposition~\cite{cowtan2020phase} (\texttt{naive}) for 5 different quantum computers.}
    \label{fig:cnot-ps}
\end{figure}

\added[id=ed]{
In \autoref{fig:cnot-ps}, the difference between the different CNOT re-synthesis strategies is shown for the ParitySynth algorithm~\cite{vandaele2022phase}. The pure ParitySynth algorithm (\texttt{ps}) generates strictly less CNOTs than the naive decomposition, but the look-ahead heuristic makes the algorithm slightly slower than naive. Re-synthesizing the trailing CNOTs is beneficial for circuits with more phase gadgets but for small circuits, the CNOT synthesis algorithm increases the number of CNOTs. This shows that the CNOT synthesis algorithms cannot find the minimal number of CNOTs needed. Using PermRowCol (\texttt{ps prc}) instead of using RowCol (\texttt{ps rc}) seems to reduce the CNOT count with an almost negligible margin. The compiler runtime increases with the complexity of the used method as expected, although the variance in runtime for PermRowCol is strange. We suspect this might be due to the way all experiments were run on the compute cluster since we do not see the variance in runtime later when PermRowCol is combined with Reverse Traversal or simulated annealing.
}

\begin{figure}
    \centering
    \includegraphics[width=0.75\linewidth]{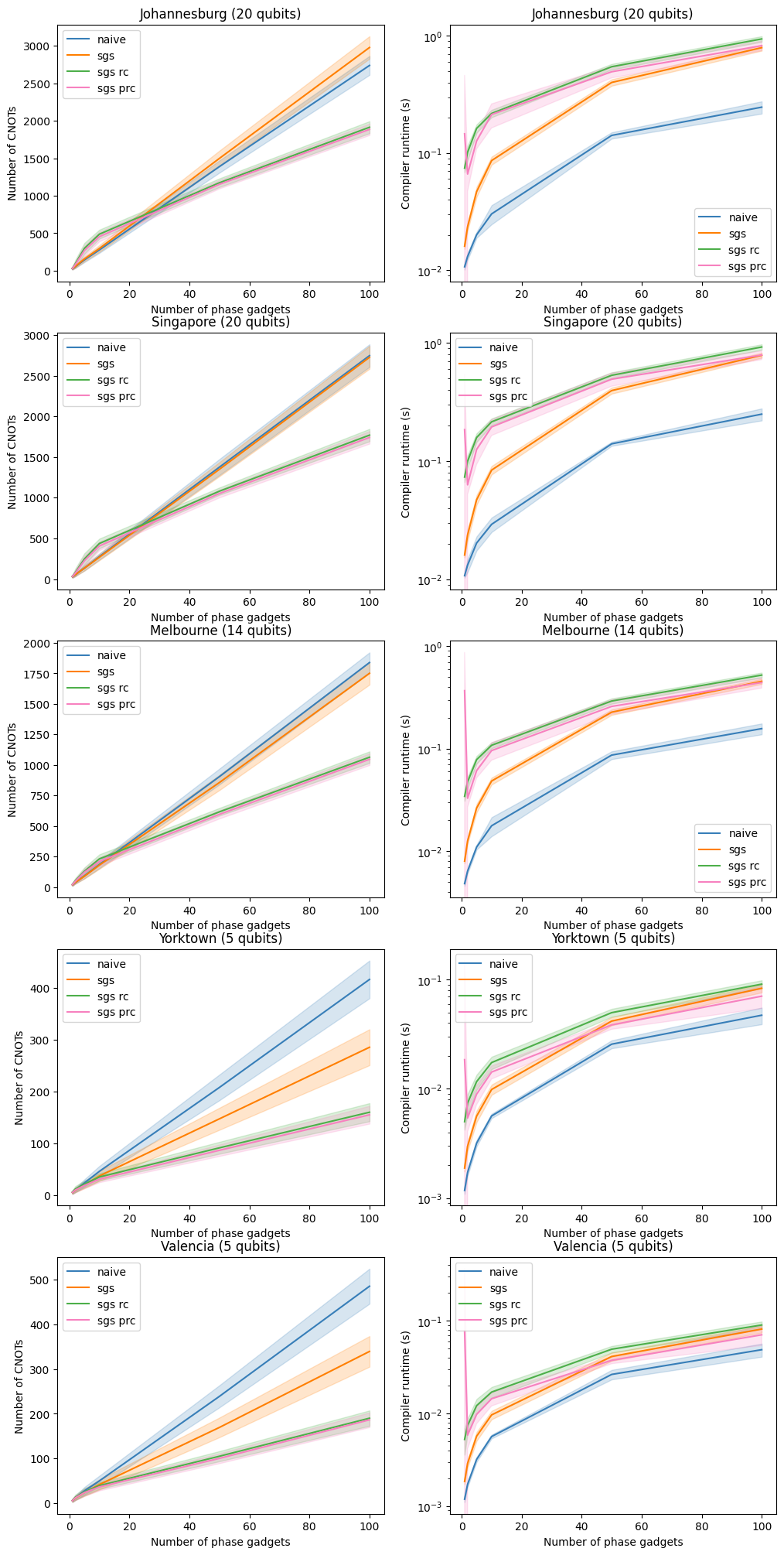}
    \caption{A comparison of CNOT count (left) and compiler runtime (right) for circuits compiled using Steiner-GraySynth~\cite{meijer-vandegriend2020architectureaware} without re-synthesizing trailing CNOTs (\textit{sgs}), with trailing CNOTs re-synthesized with RowCol~\cite{wu2019optimization} (\texttt{sgs rc}) or PermRowCol~\cite{meijer2022dynamic} (\texttt{sgs prc}), or compiled using naive decomposition~\cite{cowtan2020phase} (\texttt{naive}) for 5 different quantum computers.}
    \label{fig:cnot-sgs}
\end{figure}

\added[id=ed]{
In \autoref{fig:cnot-sgs}, the difference between the different CNOT re-synthesis strategies is shown for the Steiner-GraySynth algorithm~\cite{meijer-vandegriend2020architectureaware}. We observe similar differences as for the ParitySynth algorithm, but with two notable exceptions: (1) if the trailing CNOTs are not re-synthesized (\texttt{sgs}), Steiner-GraySynth generates more CNOTs than with naive synthesis for larger devices, and (2) the compiler runtime for PermRowCol (\texttt{sgs prc}) is more stable and, unlike ParitySynth, slightly faster than RowCol. Note that for small devices, the compiler runtime for PermRowCol is even faster than the Steiner-GraySynth without re-synthesis of the trailing CNOTs. This can happen because all circuits were compiled holistically.
}
\begin{center}
    
\fbox{\begin{minipage}{.9\textwidth}
\textbf{RQ1: Does the re-synthesis of trailing CNOTs improve performance?}

\textit{Answer:} Yes, if the circuit is large enough, at the cost of a small increase of compiler runtime.
\end{minipage}}
\end{center}


\subsection{The effect of Reverse Traversal}
\begin{figure}
    \centering
    \includegraphics[width=0.75\linewidth]{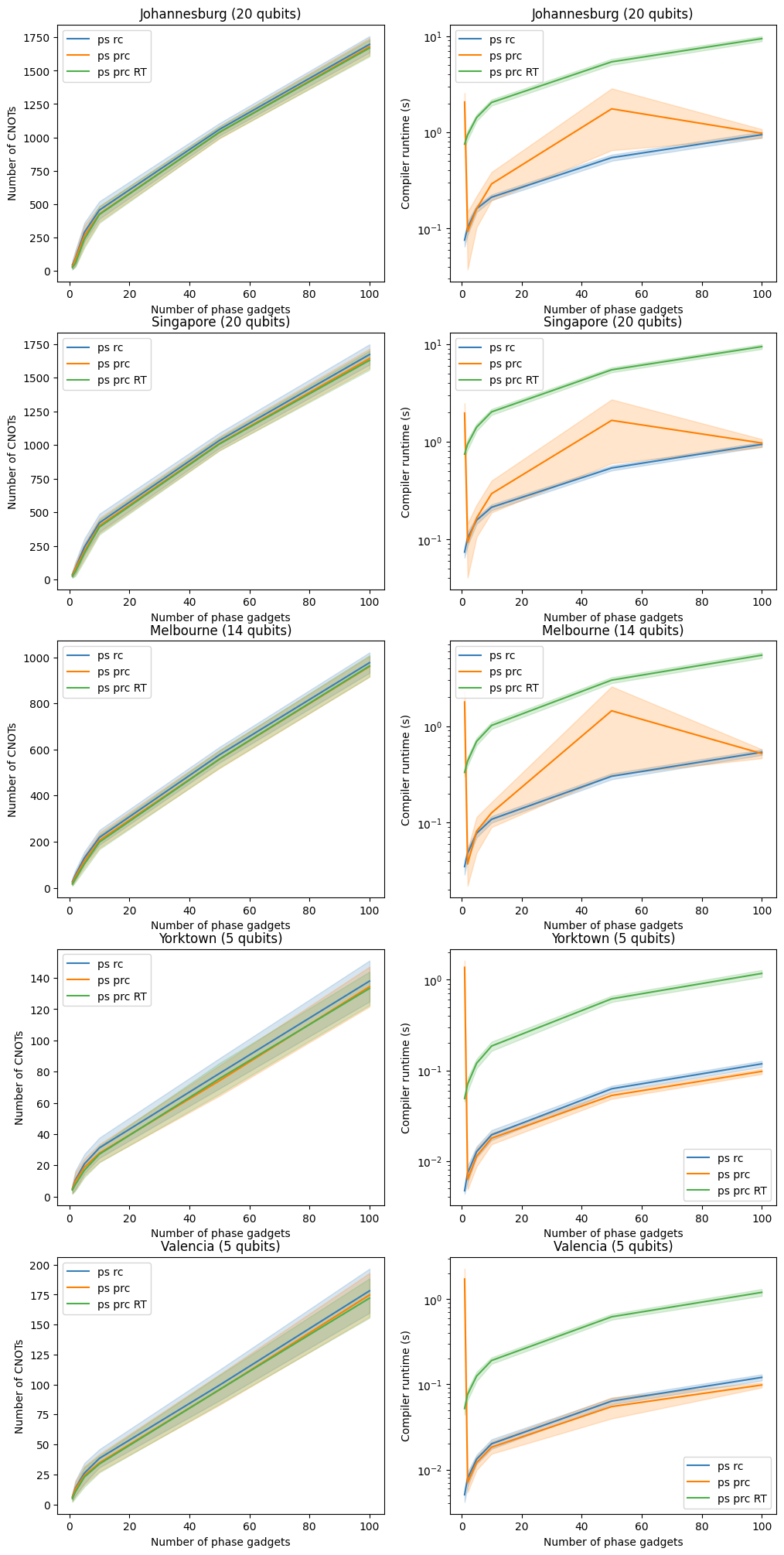}
    \caption{A comparison of CNOT count (left) and compiler runtime (right) for circuits compiled using ParitySynth~\cite{vandaele2022phase} when re-synthesizing trailing CNOTs with RowCol~\cite{wu2019optimization} (\texttt{ps rc}) or PermRowCol~\cite{meijer2022dynamic} (\texttt{ps prc}), and when using the Reverse Traversal strategy~\cite{li2019tackling} with PermRowCol (\texttt{ps prc RT}) for 5 different quantum computers.}
    \label{fig:RT-ps}
\end{figure}

\begin{figure}
    \centering
    \includegraphics[width=0.75\linewidth]{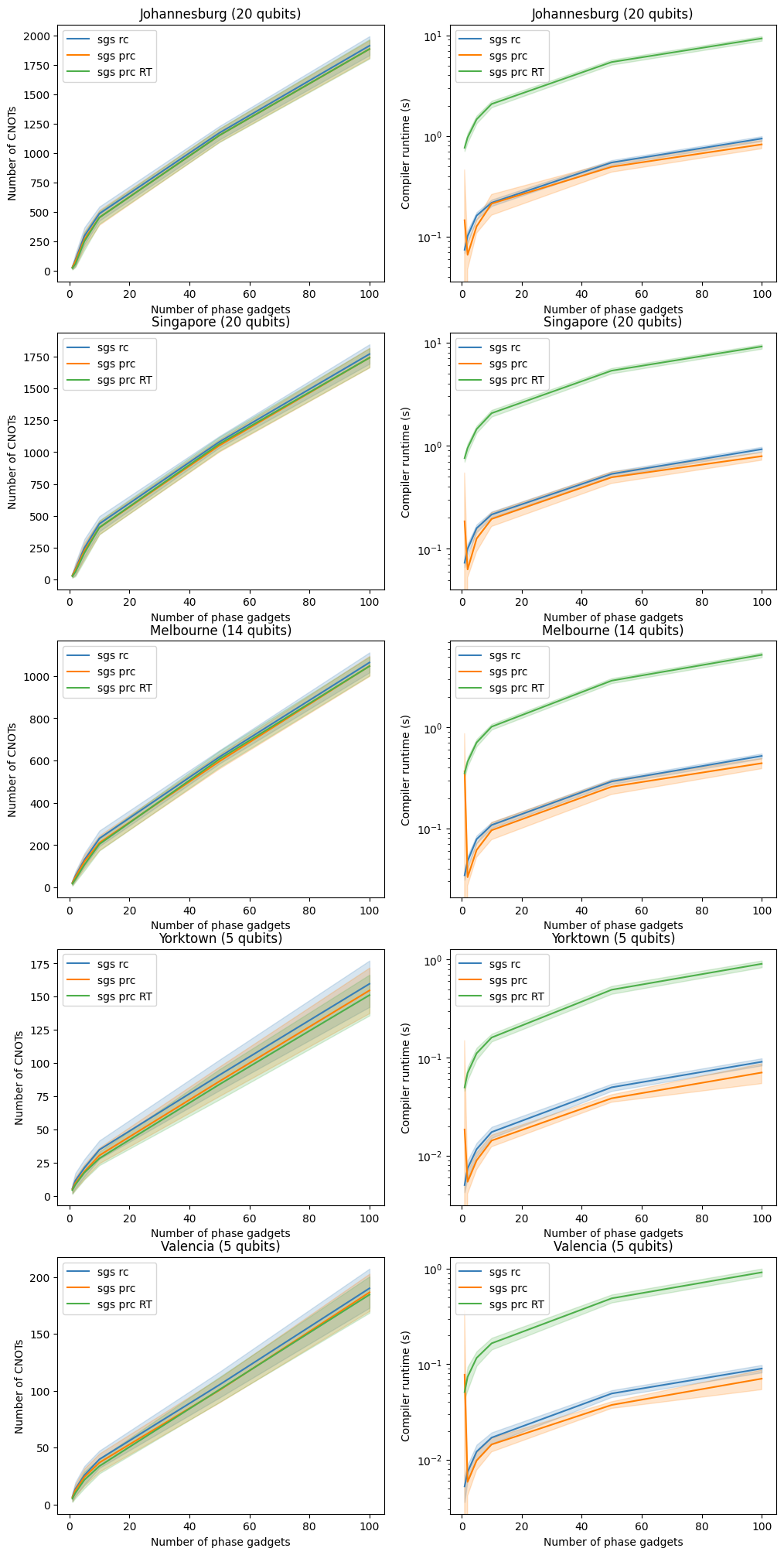}
    \caption{A comparison of CNOT count (left) and compiler runtime (right) for circuits compiled using Steiner-GraySynth~\cite{meijer-vandegriend2020architectureaware} when re-synthesizing trailing CNOTs with RowCol~\cite{wu2019optimization} (\texttt{sgs rc}) or PermRowCol~\cite{meijer2022dynamic} (\texttt{sgs prc}), and when using the Reverse Traversal strategy~\cite{li2019tackling} with PermRowCol (\texttt{sgs prc RT}) for 5 different quantum computers.}
    \label{fig:RT-sgs}
\end{figure}

\added[id=ed]{
To investigate the effect of Reverse Traversal, we plot the performance of the synthesis of the trailing CNOTs with RowCol (\texttt{rc}), PermRowCol (\texttt{prc}), or Reverse Traversal (\texttt{prc RT}) for both ParitySynth (in \autoref{fig:RT-ps}) and Steiner-GraySynth (in \autoref{fig:RT-sgs}). Note that Reverse Traversal can only be applied when the CNOTs are re-synthesized with PermRowCol. The addition of RowCol in the figures is to indicate the relative difference in CNOT count between these methods.}

\added[id=ed]{
We can see in both \autoref{fig:RT-ps} and \ref{fig:RT-sgs} that the addition of Reverse Traversal does not improve the CNOT count with respect to PermRowCol, but the plotted lines are distinguishable from RowCol re-synthesis. This is surprising because using Reverse Traversal on CNOT circuits does give smaller circuits than only using PermRowCol~\cite{meijer2022dynamic}.  Additionally, we note that Reverse Traversal increases the compiler runtime by approximately $10$ times, this is because Reverse Traversal compiles each circuit $10$ times.
}

\begin{center}
    
\fbox{\begin{minipage}{.9\textwidth}
\textbf{RQ2: Does Reverse Traversal improve performance?}

\textit{Answer:} Marginally, but it might not be worth the additional compiler runtime.
\end{minipage}}
\end{center}

\subsection{The effect of simulated annealing}
\added[id=ed]{
To investigate the effect of simulated annealing, we compare the methods from \autoref{subsec:cnot} combined with simulated annealing to the simple phase polynomial synthesis with and without the trailing CNOTs re-synthesized with PermRowCol. }

\begin{figure}
    \centering
    \includegraphics[width=0.75\linewidth]{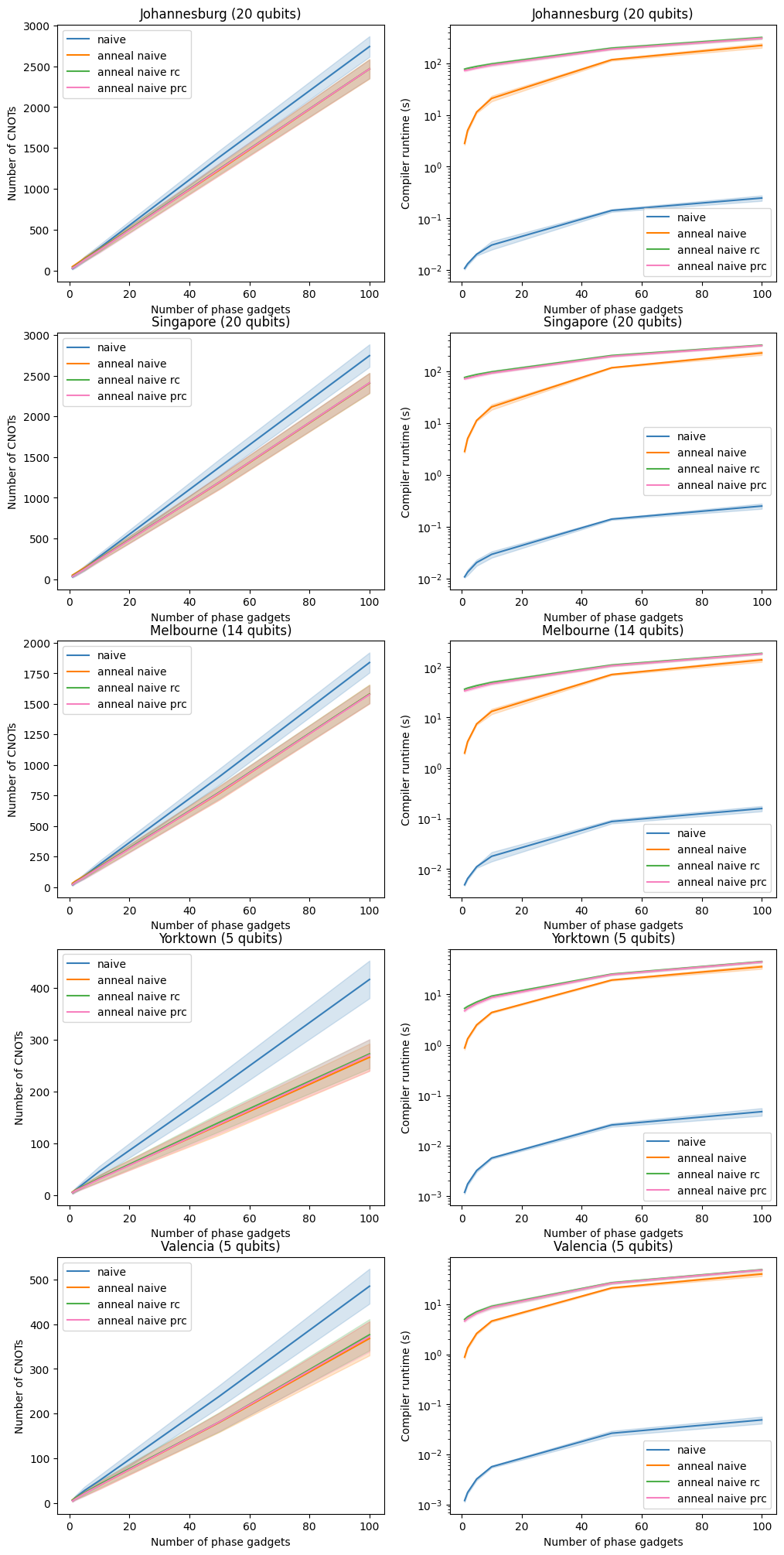}
    \caption{A comparison of CNOT count (left) and compiler runtime (right) for circuits compiled using naive decomposition~\cite{cowtan2020phase} (\texttt{naive}), or when combined with simulated annealing~\cite{gogioso2022annealing} leaving the trailing CNOTs as is (\texttt{anneal naive}), or re-synthesizing them with RowCol (\texttt{anneal naive rc}), or PermRowCol (\texttt{anneal naive prc}) for 5 different quantum computers.}
    \label{fig:anneal-naive}
\end{figure}

\added[id=ed]{
In \autoref{fig:anneal-naive}, we show the results for naive phase polynomial decomposition. Here, the RowCol and PermRowCol re-synthesis is applied to the last CNOT ladder from the phase polynomial and the CNOTs generated during annealing. Thus, there can be a small improvement from no re-synthesis, to RowCol re-synthesis, to PermRowCol, but that marginal improvement is only seen for the 5 qubit devices. For larger devices, it did not matter whether the CNOTs were re-synthesized or not. What is visible is that the annealing improves the naive decomposition significantly, as was previously shown by \citet{gogioso2022annealing}. This can be worth the increase in compiler runtime from recompiling the circuit in every iteration.}

\begin{figure}
    \centering
    \includegraphics[width=0.75\linewidth]{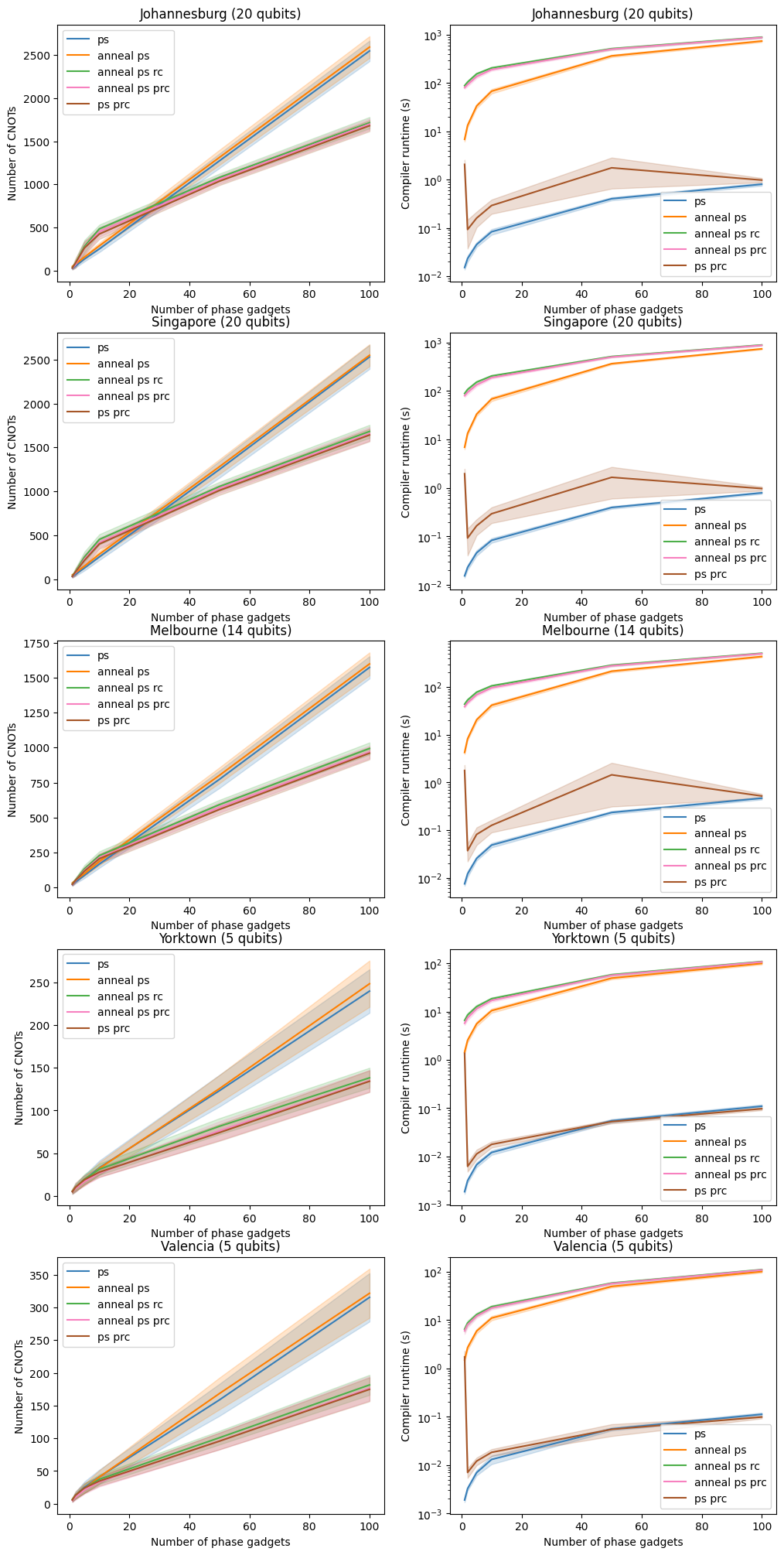}
    \caption{A comparison of CNOT count (left) and compiler runtime (right) for circuits compiled using ParitySynth~\cite{vandaele2022phase} without re-synthesizing trailing CNOTs (\texttt{ps}), or re-synthesized with PermRowCol~\cite{meijer2022dynamic} (\texttt{ps prc}), and when combined with simulated annealing~\cite{gogioso2022annealing} leaving the trailing CNOTs as is (\texttt{anneal ps}), or re-synthesizing them with RowCol (\texttt{anneal ps rc}), or PermRowCol (\texttt{anneal ps prc}) for 5 different quantum computers.}
    \label{fig:anneal-ps}
\end{figure}

\begin{figure}
    \centering
    \includegraphics[width=0.75\linewidth]{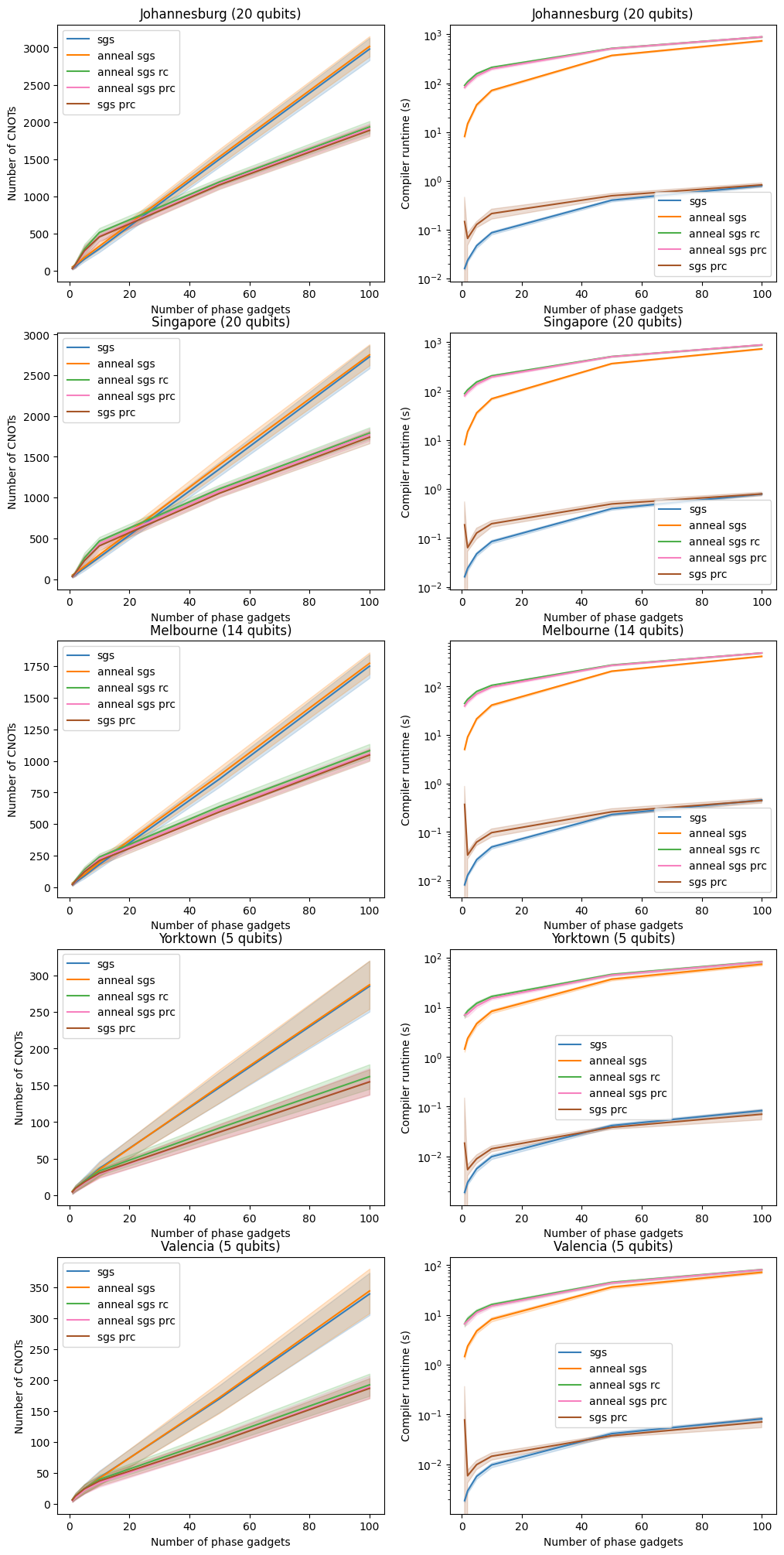}
    \caption{A comparison of CNOT count (left) and compiler runtime (right) for circuits compiled using Steiner-GraySynth~\cite{meijer-vandegriend2020architectureaware} without re-synthesizing trailing CNOTs (\texttt{sgs}), or re-synthesized with PermRowCol~\cite{meijer2022dynamic} (\texttt{sgs prc}), and when combined with simulated annealing~\cite{gogioso2022annealing} leaving the trailing CNOTs as is (\texttt{anneal sgs}), or re-synthesizing them with RowCol (\texttt{anneal sgs rc}), or PermRowCol (\texttt{anneal sgs prc}) for 5 different quantum computers.}
    \label{fig:anneal-sgs}
\end{figure}

\added[id=ed]{
On the contrary, we can see in \autoref{fig:anneal-ps} and \ref{fig:anneal-sgs} that annealing does not improve ParitySynth and Steiner-GraySynth, respectively. The annealing worsens the CNOT count in almost all cases. This shows that these phase polynomial synthesis methods already make use of common entangling structures in the phase polynomial and that cannot be improved with simulated annealing.
}

\begin{center}
    
\fbox{\begin{minipage}{.9\textwidth}
\textbf{RQ3: Does simulated annealing improve performance?}

\textit{Answer:} Only for naive decomposition.
\end{minipage}}
\end{center}


\subsection{The effect of combining Reverse Traversal with simulated annealing}
\added[id=ed]{
In the previous sections, we have seen that Reverse Traversal and simulated annealing do not improve compiler performance when combined with ParitySynth or Steiner-GraySynth. But they might have an effect when used together. Thus, we compare the best option from \autoref{subsec:cnot} to the case of running the annealer for every iteration of Reverse Traversal (\texttt{anneal and RT}, running Reverse Traversal followed by the annealer (\texttt{RT then anneal}), and iteratively rerunning the annealer and taking the best result (\texttt{iter anneal}). }

\begin{figure}
    \centering
    \includegraphics[width=0.75\linewidth]{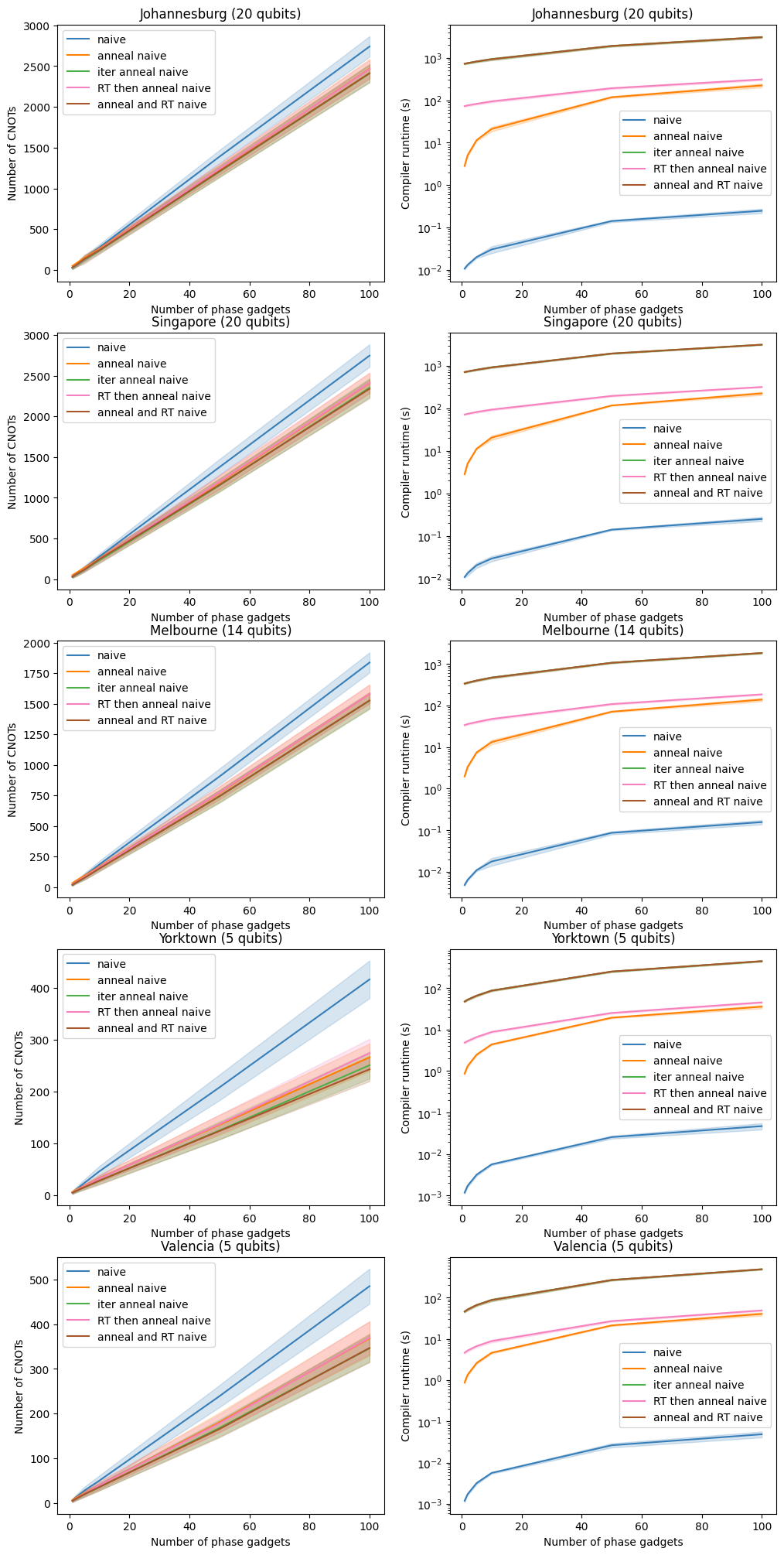}
    \caption{A comparison of CNOT count (left) and compiler runtime (right) for circuits compiled using naive decomposition~\cite{cowtan2020phase} (\texttt{naive}), when combined with simulated annealing~\cite{gogioso2022annealing}, iteratively rerunning the annealer (\texttt{iter anneal naive}), running Reverse Traversal~\cite{gogioso2022annealing} followed by the annealer (\texttt{RT then anneal naive}), or running the annealer at every step of the Reverse Traversal (\texttt{anneal and RT naive}) for 5 different quantum computers.}
    \label{fig:combined-naive}
\end{figure}

\added[id=ed]{
In \autoref{fig:combined-naive}, we see how these methods perform in combination with naive synthesis. First, it is noted that because the PermRowCol does not do much in combination with naive decomposition and annealing (see \autoref{fig:anneal-naive}), it is expected that running the annealer for every iteration of reverse traversal (\texttt{anneal and RT naive}) to result in a similar CNOT count as restarting the annealer multiple times (\texttt{iter anneal naive}). We can mostly see this in the figure, except for the Yorktown device, where the Reverse Traversal can improve. This could be because the connectivity graph is denser than that of Valencia. Similarly, there seems to be no difference between annealing and Reverse Traversal before annealing except for Yorktown. Lastly, restarting the annealer multiple times, with or without Reverse Traversal, improves the results over running the annealer once. This is because restarting helps the random nature of the annealer to find a slightly better seed.}

\begin{figure}
    \centering
    \includegraphics[width=0.75\linewidth]{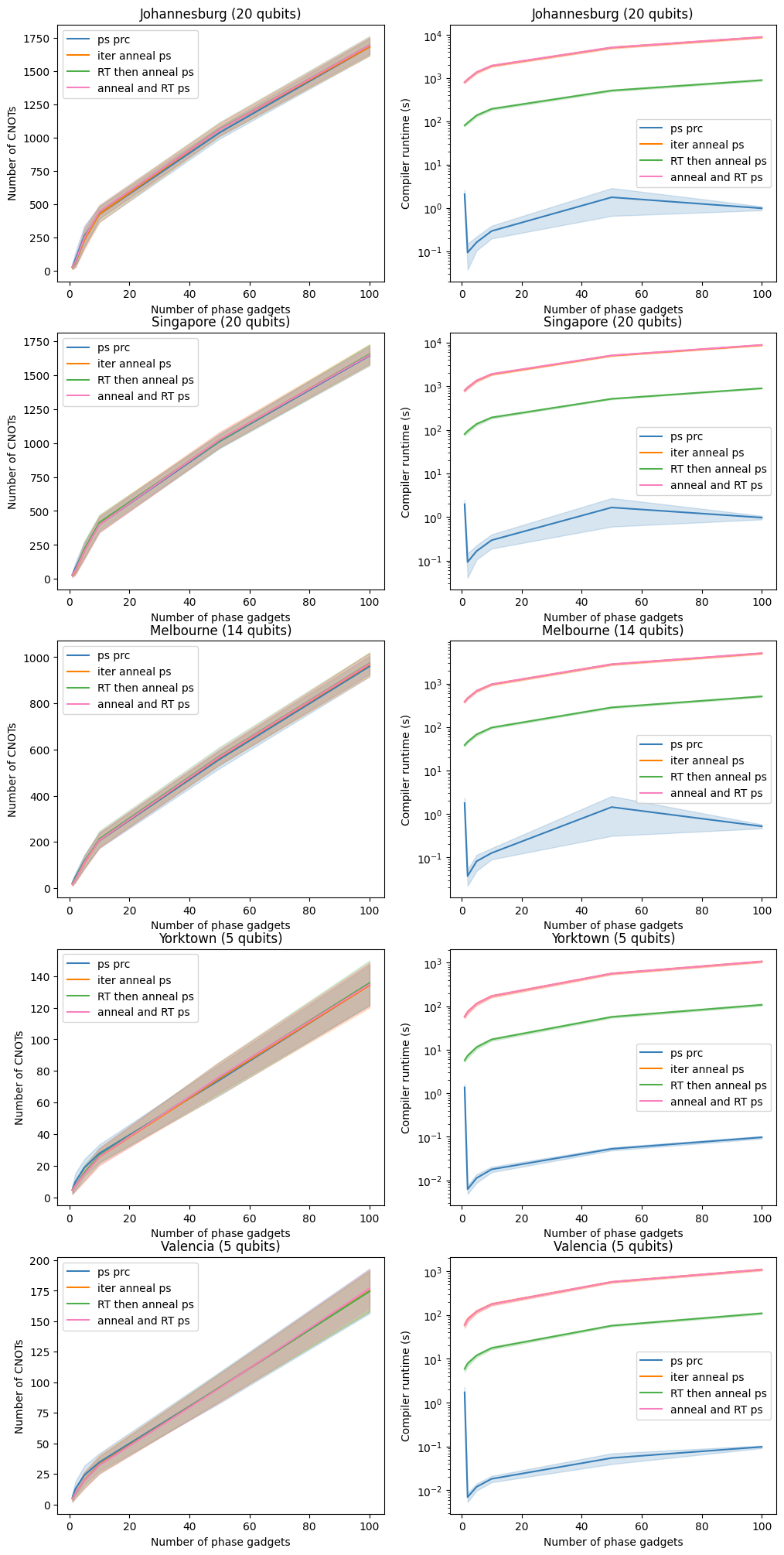}
    \caption{A comparison of CNOT count (left) and compiler runtime (right) for circuits compiled using ParitySynth~\cite{vandaele2022phase} with trailing CNOTs re-synthesized with PermRowCol~\cite{meijer2022dynamic} (\texttt{ps prc}) iteratively rerunning the annealer~\cite{gogioso2022annealing} (\texttt{iter anneal ps}), running Reverse Traversal~\cite{gogioso2022annealing} followed by the annealer (\texttt{RT then anneal ps}), or running the annealer at every step of the Reverse Traversal (\texttt{anneal and RT ps}) for 5 different quantum computers.}
    \label{fig:combined-ps}
\end{figure}

\begin{figure}
    \centering
    \includegraphics[width=0.75\linewidth]{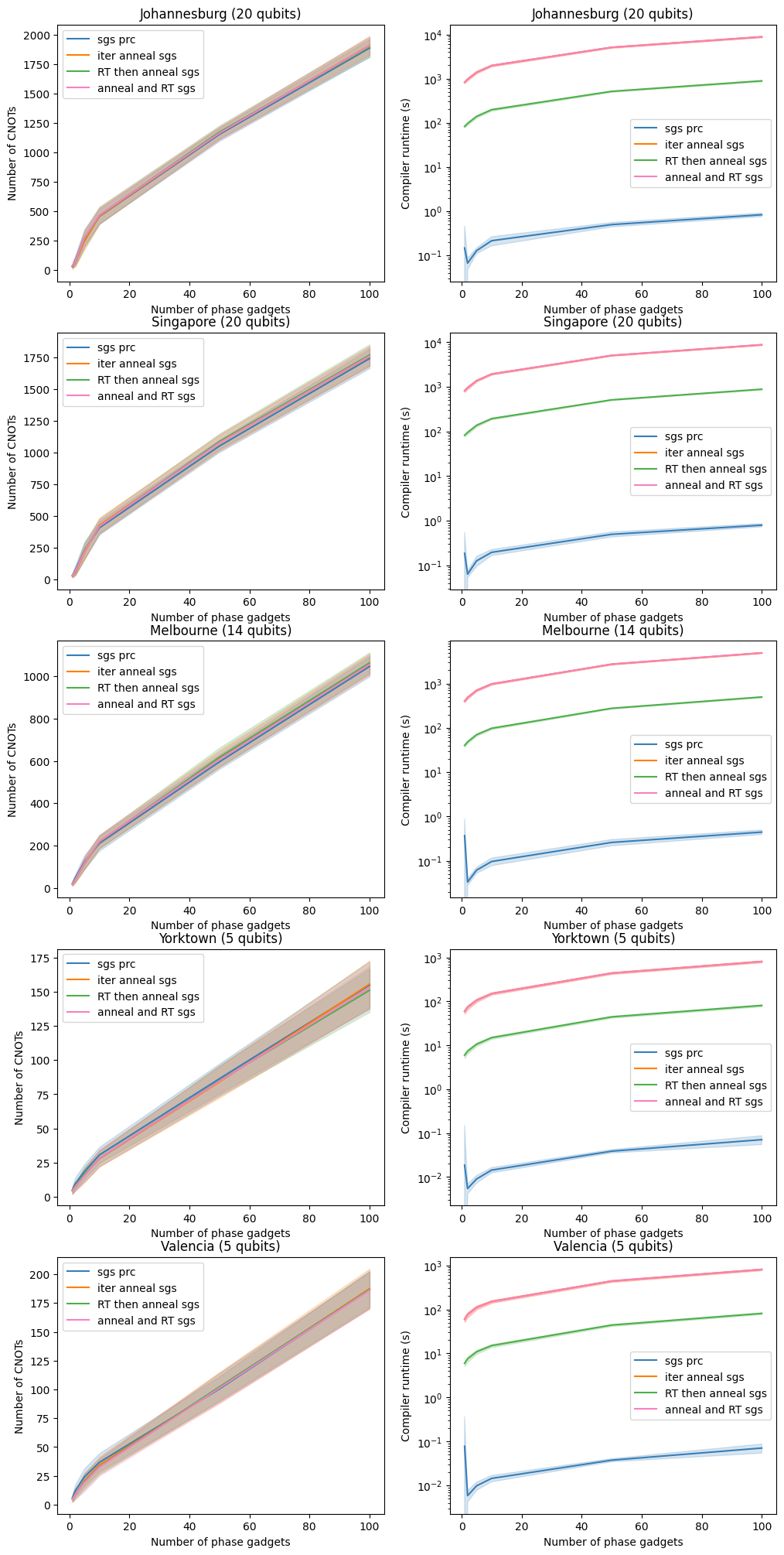}
    \caption{A comparison of CNOT count (left) and compiler runtime (right) for circuits compiled using Steiner-GraySynth~\cite{meijer-vandegriend2020architectureaware} with trailing CNOTs re-synthesized with PermRowCol~\cite{meijer2022dynamic} (\texttt{sgs prc}) iteratively rerunning the annealer~\cite{gogioso2022annealing} (\texttt{iter anneal sgs}), running Reverse Traversal~\cite{gogioso2022annealing} followed by the annealer (\texttt{RT then anneal sgs}), or running the annealer at every step of the Reverse Traversal (\texttt{anneal and RT sgs}) for 5 different quantum computers.}
    \label{fig:combined-sgs}
\end{figure}

\added[id=ed]{
On the contrary, \autoref{fig:combined-ps} and \ref{fig:combined-sgs} show that the combination of Reverse Traversal and annealing has almost no effect on the circuit size. There seem to be some small differences, but there seems to be no pattern and the differences are so small that they are most likely due to the small amount of tested circuits. If there is a minute effect, then the additional compiler runtime is probably not worth it.
}

\begin{center}
    
\fbox{\begin{minipage}{.9\textwidth}
\textbf{RQ4: Does a combination of Reverse Traversal and simulated annealing improve
performance?}

\textit{Answer:} No.
\end{minipage}}
\end{center}


\subsection{Comparison DAG-based versus phase polynomial-based compiling}
\added[id=ed]{
Lastly, we combine the findings from the previous section to determine the best phase polynomial-based compiling strategy (\texttt{best}) and compare it the the DAG-based compilers: Qiskit and TKET.}

\added[id=ed]{
For naive decomposition, we found that iterative annealing and annealing during each step of Reverse Traversal result in a similar CNOT count in a similar compiler runtime. For Yorktown, Reverse Traversal seems to be slightly better, so the recommendation for best settings when using naive decomposition is to run Reverse Traversal with simulated annealing during each iteration.}

\added[id=ed]{
For ParitySynth and Steiner-GraySynth, we have two cases. If the circuit is short, it is best to generate the phase polynomial and leave the trailing CNOTs as is. If the circuit is long, it is good to re-synthesize the trailing CNOTs using PermRowCol. Using Reverse Traversal in combination with PermRowCol only marginally improves the CNOT count, but if compiler runtime is not a limiting factor, it can be worthwhile to use. Thus, the recommendation for best settings for ParitySynth and Steiner-GraySynth is to either leave the trailing CNOTs as is or re-synthesize them with PermRowCol and Reverse Traversal if runtime allows depending on which option results in a smaller circuit. Although it is possible to keep track of the shortest circuit before and after re-synthesis of the trailing CNOTs during Reverse Traversal, the strategy for choosing the smallest circuit has been implemented here as running two compiler strategies in sequence and taking the smallest. Thus, the compiler runtime shown is the sum of these two strategies.}

\begin{figure}
    \centering
    \includegraphics[width=0.75\linewidth]{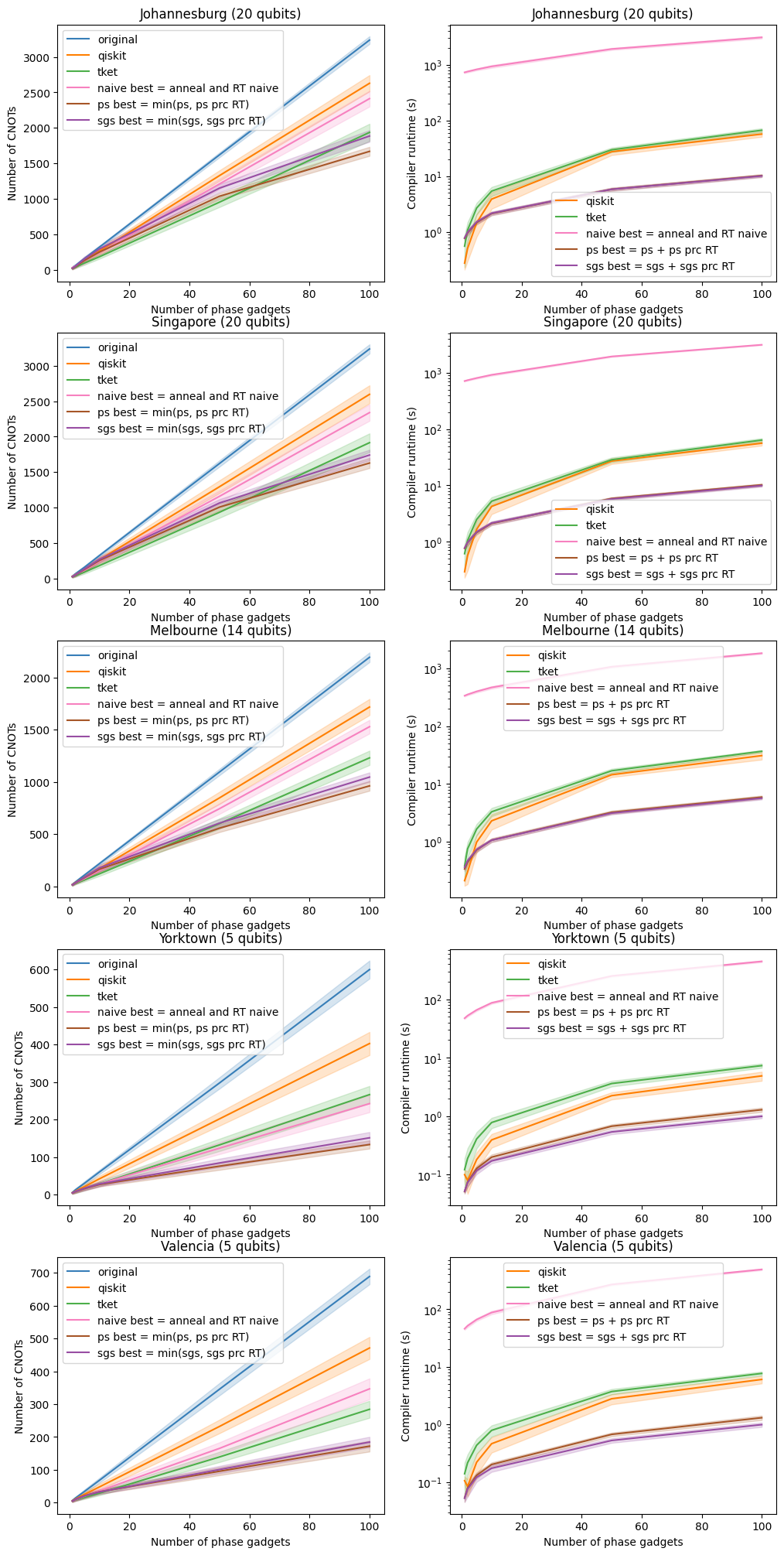}
    \caption{A comparison of CNOT count (left) and compiler runtime (right) for circuits compiled using Qiskit~\cite{Qiskit} (\texttt{qiskit}), TKET~\cite{tket} (\texttt{tket}), and the proposed best strategies for phase polynomial-based compilation using naive decomposition~\cite{cowtan2020phase} (\texttt{naive best}), ParitySynth~\cite{vandaele2022phase} (\texttt{ps best}), and Steiner-GraySynth~\cite{meijer-vandegriend2020architectureaware} (\texttt{sgs best}), as well as the original CNOT count (\texttt{original}) for 5 different quantum computers.}
    \label{fig:compare}
\end{figure}

\added[id=ed]{
The differences in CNOT count and compiler runtime for the DAG-based and phase-polynomial-based compilers are shown in \autoref{fig:compare}. First, we note that all strategies optimize the input circuit (\texttt{original}). Qiskit generates the most CNOTs, followed by the best naive synthesis strategy with the exception of the Yorktown device where the best naive synthesis is better than the TKET compiler. For small circuits, TKET is the best compiler, but for larger circuits, ParitySynth and Steiner-GraySynth generated fewer CNOTs. We can see the phase polynomial synthesis without re-synthesis of the trailing CNOTs is slightly worse than TKET, but once the re-synthesis starts to improve the CNOT count (i.e. at the bend in the lines in the plots), the additional optimization quickly catches up to TKET's performance. Additionally, ParitySynth generates strictly fewer CNOTs than Steiner-GraySynth. This is because of the additional look-ahead scheme and conforms to the findings of \citet{vandaele2022phase}. }

\added[id=ed]{
The look-ahead scheme increases the compiler runtime slightly with respect to Steiner-GraySynth but is negligible for larger devices. ParitySynth and Steiner-GraySynth are an order of magnitude faster than Qiskit and TKET, even though the additional iterations of Reverse Traversal increased the normal runtime by $10$ fold. Qiskit and TKET have a similar runtime, where TKET is slightly slower with much better CNOT counts. Lastly, the iterative nature of simulated annealing and Reverse Traversal increases the normal runtime of naive decomposition by $1000\cdot 10$ times, making it the slowest method in the figure.
}

\deleted[id=ed]{The average CNOT counts for our experiments can be found in Figure 1, Figure 2, and Figure 3, and their runtimes can be found in Figure 4. The shaded area surrounding the lines corresponds to a standard deviation from the mean of the results to indicate fluctuations.}

\deleted[id=ed]{
From Figure 1, we can see that all methods manage to optimize the given circuit. In most cases, TKET has the least CNOTs, with the exception of the Yorktown architecture where ParitySynth is better. The most CNOTs are generated by Steiner-GraySynth ($5$ qubit devices) or Qiskit (other devices). Additionally, we see that ParitySynth consistently generates fewer CNOTs than Steiner-GraySynth, which is in line with results from Vandaele et al. \[24\]. The annealer seems to get better with the size of the quantum computer and starts to improve on ParitySynth at 20 qubits. With the exception of Valencia, the annealer is better than Steiner-GraySynth.
}

\deleted[id=ed]{
From \autoref{fig:synth}, we can see that for small circuits, with a low number of phase gadgets, it is \replaced[id=1]{not beneficial to}{beneficial to not} re-synthesize the trailing CNOTs from the \deleted[id=1]{phase} phase polynomial. However, for larger circuits, the re-synthesis can reduce the number of generated CNOTs dramatically, eventually even improving with respect to the TKET compiler. We also see that although the difference between the use of RowCol, PermRowCol, or PermRowCol with Reverse Traversal is noticeable (each respectively generating fewer CNOTs), the CNOT reduction is rather small with respect to the standard deviation. Particularly when comparing smaller circuits, where these methods do not work well. This shows that neither the phase polynomial synthesis algorithms nor the CNOT synthesis algorithms \replaced[id=me]{do not}{do} find the optimal solution.
}

\deleted[id=ed]{
One can attempt to improve the synthesis methods with supplementary algorithms \replaced[id=1]{such as}{like} Reverse Traversal~\cite{li2019tackling} or simulated annealing (\replaced[id=1]{e.g.}{like in} \cite{gogioso2022annealing}\added[id=me]{)}. In \autoref{fig:anneal}, we show how these algorithms improve the results. We see that these algorithms can help in the different situations investigated, but we also note that these differences are marginal with respect to the standard deviations. We also see that the annealer struggles to improve with respect to Reverse Traversal when using a phase polynomial synthesis algorithm, rather than naive synthesis. This might be due to the low number of iterations, but it could also indicate that the optimizations found by the annealer are also those leveraged by the phase polynomial synthesis algorithms. More research is needed to distinguish these cases, but it should be clear that these supplementary algorithms can only find minor reductions in CNOT counts and large improvements might only be found by improving the synthesis algorithms themselves.
}

\deleted[id=ed]{
Lastly, we compare the runtimes of our different experiments in Figure 4. We can see in Figure 4a that the baseline synthesis algorithms are very fast and the annealer is fairly slow because it needs to resynthesize the circuit for many iterations. The Qiskit transpiler and TKET compiler both lie in between these strategies with respect to runtime. Looking at Figure 4b, we can see that naive synthesis is the fastest. The runtime of ParitySynth and Steiner-GraySynth are surprisingly similar.The addition of RowCol increases the runtime, as expected. What was unexpected is the volatile runtime of PermRowCol. For Steiner-GraySynth, it is slightly faster than RowCol, which might be because the dynamic allocation makes the Steiner Trees slightly smaller and thus the traversal faster. However, for ParitySynth, the runtime is very inconsistent with a large standard deviation. We expect this to be an artifact of the HPC cluster on which these results were run in parallel. Looking at Figure 4c, we see that the choice of supplementary algorithm can have a significant toll on the runtime of the algorithm. This is because of the number of iterations required for each algorithm the annealer or Reverse Traversal to converge. Thus, we see that using the annealer can take several minutes to compile. If the annealer is combined with the additional iterations from Reverse Traversal (or iterative annealing), it can even take $2$ hours to compile a single $100$ phase gadget circuit on a $20$ qubit device. Although this runtime could have been improved with dynamic programming, as was used by \citet{gogioso2022annealing}, we do not find this an acceptable runtime given that we also need to increase the number of iterations of the annealer to improve the CNOT count beyond the performance of the Reverse Traversal alone. Additionally, the required number of iterations should also scale with the size of the quantum computer, so this strategy might not scale for future devices. Nevertheless, if the small improvement obtained by the annealer is enough to make the resulting quantum circuit executable on a quantum device, it might be worthwhile to wait that long.
}

\begin{center}
    
\fbox{\begin{minipage}{.9\textwidth}
\textbf{RQ5: What is the performance difference between DAG-based compiling and
phase polynomial-based compiling?}

\textit{Answer:} For deep quantum circuits, phase polynomial-based compiling is much faster than DAG-based compiling and results in fewer CNOT gates. 
\end{minipage}}
\end{center}

\section{Discussion}\label{sec:discussion}
In this section, we discuss various assumptions that are made when using architecture-aware synthesis \replaced[id=me]{in phase polynomial-based}{for quantum} compiling. Additionally, these topics might provide further research directions for the continued improvement of these algorithms.

\subsection{Original circuit quality}
The architecture-aware synthesis approach to compilation assumes that the original creator of the quantum circuit to be compiled is not familiar with quantum circuit optimization and the qubit routing problem. Therefore, we assume that the given circuit is poorly optimized. 
This is a sensible assumption because the current use cases for quantum computers are the simulation of quantum processes. Thus, the users of quantum computers are physicists and chemists, who might not be familiar with quantum computer science. Similarly, computer scientists who use quantum machine learning treat the quantum circuit as a black box to be trained, similar to neural networks, and thus, they are also unfamiliar with circuit optimization and compiling.

As such, we expect that most NISQ programs will be generated from a given Hamiltonian.
During this process, the Hamiltonian is transformed into a sequence of Paulistrings, which makes the mixed ZX-phase polynomial a native intermediate representation to the original program. In fact, using Paulistrings as an intermediate representation was already proposed for the Paulihedral compiler~\cite{li2022paulihedral}.

We mention here that if the original quantum circuit is already heavily optimized for the target hardware, we do not expect these synthesis methods to improve the circuit. 
As we already could see in the results for small circuits with few phase gadgets.

\subsection{The native multi-qubit interactions of the quantum computer}
The only multi-qubit interaction that is generated by \replaced[id=me]{phase polynomial-based compilers}{our algorithm} is the CNOT gate. 
This makes sense from the point of view that together with arbitrary single qubit gates, this makes a universal gate set.
However, the physical implementation of native multi-qubit interactions on current quantum hardware is generally not a CNOT. The devices can execute a CNOT using their native interaction and some single qubit gates, but it is not immediately a CNOT. Some compilation is still necessary!

In general, it takes one of these multiqubit interactions to simulate the CNOT, making it seem that this does not affect the number of multiqubit interactions. However, some devices (e.g. ion-traps) allow multi-qubit interactions between all qubits. This would remove the need to reduce the phase gadgets into smaller interactions. 

But let's assume, for the sake of argument, that we have a device with a native 2-qubit interaction. In that case, we can still simulate the CNOT using some single-qubit gates. 
However, on some devices these interactions will be implemented in an ``echoed`` form to \replaced[id=1]{mitigate}{improve} environmental noise~\cite{earnest2021pulseefficient}, doubling all 2-qubit gates. Moreover, in cases where a 2-qubit phase gadget acts on two connected qubits on the topology, the phase gadget might be turned into a native 2-qubit interaction directly (using single-qubit gates) without the need for generating CNOTs.

Compiling a phase gadget directly to the native 2-qubit gate would require that 2-qubit gate to be tunable with the angle of the phase gadget. Some devices allow this (e.g. some IBM devices~\cite{earnest2021pulseefficient}), in which case they need to be calibrated for all possible angles rather than just the one angle needed for the CNOT (or two if it is echoed). But calibration of many angles is more difficult. So herein lies a trade-off: implement a 2-qubit phase gadget with 2 ``good`` interactions (from CNOTs) or 1 ``bad`` interaction (directly native)? 

The architecture-aware synthesis procedures will change the balance of this trade-off. 
When synthesizing the last CNOT of a phase gadget, we can decide to either synthesize the native interaction instead of the CNOT or synthesize the CNOT and push it through the remainder of the intermediate representation such that it might be optimized away. Then the intermediate representation of the remaining circuit can be very different depending on the choice. Continuing the synthesis procedure, each choice between CNOT or native gate will influence the final CNOT count. However, it is not obvious when each option is better because we don't know if a CNOT will be optimized away or make the remaining phase gadgets more difficult to synthesize.

What is clear in this trade-off is that, because the second interaction might be optimized away, the quality of the tunable 2-qubit interaction needs to be better than twice the fixed-angle interaction. 

Moreover, these architecture-aware synthesis approaches are not suitable for a look-up-and-replace strategy to introduce the tunable 2-qubit interactions, because they probably will not generate CNOT-(single qubit gate)-CNOT sequences. Thus, in case tunable interactions are allowed, the synthesis approach will need to take that into account from the start. 

\subsection{The native single qubit gates}
The only single-qubit gates that these architecture-aware synthesis approaches generate are X and Z rotations. The assumption here is that these are equally desirable for the hardware, but this is not necessarily true.
For example, in superconducting devices, rotations around the Z axis are ``virtual``. They do not change the qubit itself but the equipment around it (i.e. it resets the clock). This means that Z rotations are essentially noiseless while X rotations require noise manipulations of the qubit. Thus, it is beneficial to generate more Z rotations than X rotations. 

Given that X rotations will have to be calibrated and Z rotations do not, it is better to calibrate the X rotations for one or two angles and simulate all arbitrary X rotations $R_X(\alpha)$ as $HR_Z(\alpha)H$, where the $H$ gate is created from the calibrated X rotations and the virtual Z rotation.

This adjustment can be immediately added to our proposed algorithm, but we can also make a slight adjustment to \replaced[id=me]{phase polynomial synthesis}{our algorithm} that might influence the design of future quantum hardware. During the synthesis procedure, we alternate between decomposing sequences of Z-phase gadgets and X-phase gadgets. What we can do is generate a $H$ gate on every qubit before the sequence of X-phase gadgets and after the sequence of X-phase gadgets. Due to the connectivity constraints, we do not know which qubits are needed to synthesize the X-phase gadgets, so we need to put the $H$ gate on all qubits (unnecessary $H$ gates can be removed after synthesis). Now, the sequence of X-phase gadgets is turned into a sequence of Z-phase gadgets, and we can continue the algorithm \replaced[id=1]{similar to}{like} before, except that when a generated CNOT is pushed through the barrier of $H$ gates, its direction is reversed. In case the sequence of X-phase gadgets is longer than the number of qubits, this reduces the total amount of $H$ gates required.

The reason we believe strategy might influence the design of future quantum hardware is that placing a $H$ gate on all qubits is a very common operation in quantum algorithms~\cite{Farhi_Goldstone_Gutmann_2014,Shor_1994,Grover_1996}. If the compilation also generates these frequently, it might be sensible to make this operation ``hardware accelerated``. I.e. make some hardware adjustments so that this specific thing can be done better, faster, and/or easier.

\subsection{Effect of gate fidelity}
Another effect that \replaced[id=me]{phase polynomial synthesis}{our algorithm} does not natively take into account is the difference in gate fidelity. On current quantum computers, not every qubit can be equally well controlled and not every multi-qubit interaction has the same noise. Our results do not take this into account and assume that all gates are equally good. We have discussed some of these aspects when discussing the implementation of the natively supported gates by the quantum devices, but we will discuss the effect of fidelity differences in a broader sense here.

Although we assume for our results that all gates are equally good, \replaced[id=me]{phase polynomial synthesis}{our algorithm} can be adjusted to take gate quality into account. The most straightforward way to do this is to make the connectivity graph a weighted graph where the weights correspond to the gate fidelities. More generally, we can redefine the measure of ``qubit distance`` when calculating the minimal Steiner trees. Within our implementation, this would mean redefining the value of the shortest path between every qubit in the topology. We approximate the Steiner tree by calculating the minimal spanning tree over the all-pairs shortest paths. This measure of qubit distance can then take more into account than only the 2-qubit gate fidelity, but also the quality of the qubit itself. Similarly, we can use gate fidelity as a tie-breaker when choosing CNOTs or qubits in \replaced[id=me]{phase polynomial synthesis}{our algorithm}. 

In general, we need to find better strategies to take more dynamic sources of noise into account. The algorithm can be adjusted to reduce crosstalk between qubits, improve gate scheduling, or take into account other device-specific sources of error. This will require more collaboration between hardware manufacturers and compiler designers, as well as the availability of more detailed specifications for the available hardware.

\subsection{Effect of approximate Steiner-trees}
Lastly, we discuss the elephant in the room: finding a minimal Steiner tree is an NP-hard problem~\cite{Steiner}. The only reason \replaced[id=me]{phase polynomial-based compiling}{our algorithm} is performant is that we use a polynomial approximation. This means that the paths that we find between qubits might not be optimal and we could reduce the CNOT count even further. 

It is possible that the types of connectivity graphs on quantum computers are sparse enough so that the approximation is able to find the minimum Steiner tree, but we do not know.

In case the approximate Steiner trees are insufficient, we could still find a minimal Steiner tree using quantum computers. Thanks to adiabatic quantum computing, we should be able to find minimal Steiner trees with quantum annealers~\cite{Lucas_2014}. In theory, we could also use gate-based quantum computers to do this, but we run into a chicken-egg problem. Because current quantum annealers have more qubits to their disposal and they do not need a transpiled quantum circuit, we can actually use quantum computers to compile quantum circuits. In fact, this has recently already been done in the ISAAQ compiler~\cite{naito2023isaaq}.

More generally, we might be able to describe the full parity network problem (solved in Paritysynth) in a fashion suitable for quantum annealers. There is a lot of interaction between generating which parity on which qubit in which order, which might be very suitable for a quantum annealing formulation.

\subsection{Paulistrings and Clifford tableau as intermediate representation}
One downside of using mixed-ZX phase polynomials as an intermediate representation is that we need to decompose Paulistrings into sequences of Z- and X-phase gadgets. This is inefficient because that will treat single-qubit Clifford gates as phase gadgets, rather than Clifford gates. For Clifford gates, we know how they commute through Paulistrings~\cite{cowtan2020phase} and we can efficiently represent them in a Clifford tableau. This Clifford tableau is very similar to a parity matrix and we could attempt to find an architecture-aware synthesis algorithm for these Clifford tableaus. In fact, such an algorithm has recently been proposed~\cite{winderl2023architectureaware}. So, if we adjust the architecture-aware phase polynomial synthesis algorithm to take arbitrary Paulistrings, we can use the same techniques described in this paper directly for trotterized Hamiltonians and hardware-efficient ansatze, such as those found in hybrid quantum computing.

\section{Conclusion and future work}\label{sec:conclusion}
\added[id=ed, comment={Rewrote conclusion section completely}]{
In conclusion, we have compared the performance of current DAG-based compilers against that of phase polynomial-based compiling strategies. In doing so, we have combined various proposed methods for phase polynomial-based compiling to determine a best practice for the use of these methods.
}

\added[id=ed]{
First, we confirmed that for phase polynomial-based compilation, ParitySynth outperforms Steiner-GraySynth, and naive decomposition results in inefficient circuits.
Additionally, we confirmed that re-synthesis of trailing CNOTs after phase polynomial synthesis can greatly reduce the number of CNOTs for large circuits at a small increase of compiler runtime. However, for shallow circuits, the CNOT synthesis is unable to optimize the number for trailing CNOTs. This indicates a need for improved CNOT circuit synthesis algorithms that are CNOT efficient for short CNOT circuits. 
}

\added[id=ed]{
We found that adding Reverse Traversal to phase polynomial-based compilation can decrease the CNOT count slightly but the required additional iterations of the compiler might not be worth the additional compiler time. This contrasts the results reported by \citet{meijer2022dynamic} for CNOT circuits. However, at the time of writing, the compiler runtime of phase polynomial-based compiling with Reverse Traversal is much shorter than the DAG-based compilers in our experiments (i.e. Qiskit and TKET). As such, the current recommendation is to use Reverse Traversal if time allows, since it cannot result in a larger CNOT count.
}

\added[id=ed]{
Additionally, we found that the addition of simulated annealing only improves the naive decomposition of phase polynomials and actually hinders the more sophisticated algorithms:  ParitySynth and Steiner-Graysynth. This result still holds when combining simulated annealing with Reverse Traversal. However, we suspect that any improvement obtained from naive synthesis with Reverse Traversal during every iteration comes from restarting the simulated annealer since it performed similar to iteratively restarting the annealer without Reverse Traversal and the PermRowCol algorithm in the Reverse Traversal loops does not have many degrees of freedom to improve CNOT count when using naive decomposition of the phase polynomials. 
}

\added[id=ed]{
Lastly, we have shown that the phase polynomial-based compiling methods outperform current popular DAG-based compilers Qiskit and TKET for very large circuits that are beyond the capabilities of current quantum computers. Moreover, it does so in a fraction of the time that is needed for TKET and Qiskit to compile the circuits. For small circuits, that are executable on current devices, Qiskit and TKET create more optimized circuits, i.e. with less CNOTs. 
}

\added[id=ed]{
As larger quantum computers are being developed and longer quantum circuits need to be executed, time-efficient and CNOT-efficient quantum compilers are soon necessary. Thus, research on the optimality of different compiler strategies is becoming more urgent. In particular, we need to find both theoretical and practical minimality results for phase polynomial synthesis both in gate count and compiler runtime, we need to find strategies to combine DAG-based methods and phase polynomial-based methods to use the best of both strategies, and we need to make these methods available to the public in a user-friendly manner.
}

\deleted[id=ed]{
In this paper, we have compared different architecture-aware synthesis strategies for compiling quantum circuits. Additionally, we have compared their performance with respect to common alternatives, such as Qiskit and TKET. We also investigated to effect of supplementary algorithms, such as Reverse Traversal and simulated annealing, on these synthesis algorithms.
}

\deleted[id=ed]{
We have shown that although the synthesis algorithms perform well for large circuits with respect to the existing compilers, they do not synthesize an optimal circuit for small circuits. Additionally, the addition of supplementary algorithms can improve the results, but they do so with a very small margin. 
}

\deleted[id=ed]{
Thus we conclude that these architecture-aware synthesis strategies are effective for the compilation of medium-scale hybrid quantum circuits because the intermediate representation of these circuits is very closely related to the problem representation of hybrid quantum algorithms. However, more research is needed to use also use these algorithms for smaller quantum circuits. 
}

\section*{Acknowledgement}
I would like to thank Richie Yeung and Stefano Gogioso for helpful discussions and insights into their code base on which the full algorithm was built. I would also like to thank Vivien Vandaele for answering my questions related to Paritysynth. Lastly, I would like to thank the anonymous reviewers for their actionable feedback.

\added[id=me]{This work was partially funded by the Business Finland Quantum Computing Campaign, project FrameQ.
}

\added[id=me, comment={Updated \cite{meijer-vandegriend2020architectureaware,meijer2022dynamic,gogioso2022annealing,tket, nash2020quantum,li2019tackling,wetering2020zx,naito2023isaaq,Lucas_2014}}]{}
\deleted[id=1, comment={Updated author reference \cite{Qiskit}}]{}
\deleted[id=2, comment={Updated author reference \cite{Qiskit}}]{}
\deleted[id=1, comment={Update title in reference \cite{wu2019optimization}}]{}

\bibliographystyle{elsarticle-harv}
\bibliography{tex}

\providecommand{\noopsort}[1]{}
\begin{thebibliography}{27}
\expandafter\ifx\csname natexlab\endcsname\relax\def\natexlab#1{#1}\fi
\providecommand{\url}[1]{\texttt{#1}}
\providecommand{\href}[2]{#2}
\providecommand{\path}[1]{#1}
\providecommand{\DOIprefix}{doi:}
\providecommand{\ArXivprefix}{arXiv:}
\providecommand{\URLprefix}{URL: }
\providecommand{\Pubmedprefix}{pmid:}
\providecommand{\doi}[1]{\href{http://dx.doi.org/#1}{\path{#1}}}
\providecommand{\Pubmed}[1]{\href{pmid:#1}{\path{#1}}}
\providecommand{\bibinfo}[2]{#2}
\ifx\xfnm\relax \def\xfnm[#1]{\unskip,\space#1}\fi
\bibitem[{{Amazon Web Services}(2023)}]{braket}
\bibinfo{author}{{Amazon Web Services}}, \bibinfo{year}{2023}.
\newblock \bibinfo{title}{{Amazon Braket}}.
\newblock \URLprefix \url{https://aws.amazon.com/braket/}.
\bibitem[{Amy et~al.(2018)Amy, Azimzadeh and Mosca}]{amy2018controlled}
\bibinfo{author}{Amy, M.}, \bibinfo{author}{Azimzadeh, P.}, \bibinfo{author}{Mosca, M.}, \bibinfo{year}{2018}.
\newblock \bibinfo{title}{On the controlled-{{NOT}} complexity of controlled-{{NOT}}\textendash phase circuits}.
\newblock \bibinfo{journal}{Quantum Science and Technology} \bibinfo{volume}{4}, \bibinfo{pages}{015002}.
\newblock \URLprefix \url{https://doi.org/10.1088/2058-9565/aad8ca}, \DOIprefix\doi{10.1088/2058-9565/aad8ca}.
\bibitem[{Amy et~al.(2014)Amy, Maslov and Mosca}]{amy2014polynomialtime}
\bibinfo{author}{Amy, M.}, \bibinfo{author}{Maslov, D.}, \bibinfo{author}{Mosca, M.}, \bibinfo{year}{2014}.
\newblock \bibinfo{title}{Polynomial-{{Time T-Depth Optimization}} of {{Clifford}}+{{T Circuits Via Matroid Partitioning}}}.
\newblock \bibinfo{journal}{IEEE Transactions on Computer-Aided Design of Integrated Circuits and Systems} \bibinfo{volume}{33}, \bibinfo{pages}{1476--1489}.
\newblock \DOIprefix\doi{10.1109/TCAD.2014.2341953}.
\bibitem[{{\noopsort{brugi{\`e}re}}{de Brugi{\`e}re} et~al.(2020){\noopsort{brugi{\`e}re}}{de Brugi{\`e}re}, Baboulin, Valiron, Martiel and Allouche}]{debrugiere2020quantum}
\bibinfo{author}{{\noopsort{brugi{\`e}re}}{de Brugi{\`e}re}, T.G.}, \bibinfo{author}{Baboulin, M.}, \bibinfo{author}{Valiron, B.}, \bibinfo{author}{Martiel, S.}, \bibinfo{author}{Allouche, C.}, \bibinfo{year}{2020}.
\newblock \bibinfo{title}{Quantum {{CNOT Circuits Synthesis}} for {{NISQ Architectures Using}} the {{ Decoding Problem}}}, in: \bibinfo{editor}{Lanese, I.}, \bibinfo{editor}{Rawski, M.} (Eds.), \bibinfo{booktitle}{Reversible {{Computation}}}. \bibinfo{publisher}{{Springer International Publishing}}, \bibinfo{address}{{Cham}}. volume \bibinfo{volume}{12227}, pp. \bibinfo{pages}{189--205}.
\newblock \URLprefix \url{http://link.springer.com/10.1007/978-3-030-52482-1_11}, \DOIprefix\doi{10.1007/978-3-030-52482-1_11}.
\bibitem[{Cowtan et~al.(2020)Cowtan, Dilkes, Duncan, Simmons and Sivarajah}]{cowtan2020phase}
\bibinfo{author}{Cowtan, A.}, \bibinfo{author}{Dilkes, S.}, \bibinfo{author}{Duncan, R.}, \bibinfo{author}{Simmons, W.}, \bibinfo{author}{Sivarajah, S.}, \bibinfo{year}{2020}.
\newblock \bibinfo{title}{Phase {{Gadget Synthesis}} for {{Shallow Circuits}}}.
\newblock \bibinfo{journal}{Electronic Proceedings in Theoretical Computer Science} \bibinfo{volume}{318}, \bibinfo{pages}{213--228}.
\newblock \URLprefix \url{http://arxiv.org/abs/1906.01734v2}, \DOIprefix\doi{10.4204/EPTCS.318.13}.
\bibitem[{Earnest et~al.(2021)Earnest, Tornow and Egger}]{earnest2021pulseefficient}
\bibinfo{author}{Earnest, N.}, \bibinfo{author}{Tornow, C.}, \bibinfo{author}{Egger, D.J.}, \bibinfo{year}{2021}.
\newblock \bibinfo{title}{Pulse-efficient circuit transpilation for quantum applications on cross-resonance-based hardware}.
\newblock \bibinfo{journal}{Physical Review Research} \bibinfo{volume}{3}, \bibinfo{pages}{043088}.
\newblock \URLprefix \url{https://link.aps.org/doi/10.1103/PhysRevResearch.3.043088}, \DOIprefix\doi{10.1103/PhysRevResearch.3.043088}.
\bibitem[{Farhi et~al.(2014)Farhi, Goldstone and Gutmann}]{Farhi_Goldstone_Gutmann_2014}
\bibinfo{author}{Farhi, E.}, \bibinfo{author}{Goldstone, J.}, \bibinfo{author}{Gutmann, S.}, \bibinfo{year}{2014}.
\newblock \bibinfo{title}{A quantum approximate optimization algorithm} \URLprefix \url{https://arxiv.org/abs/1411.4028}, \DOIprefix\doi{10.48550/ARXIV.1411.4028}.
\bibitem[{Gogioso and Yeung(2023)}]{gogioso2022annealing}
\bibinfo{author}{Gogioso, S.}, \bibinfo{author}{Yeung, R.}, \bibinfo{year}{2023}.
\newblock \bibinfo{title}{Annealing optimisation of mixed zx phase circuits}.
\newblock \bibinfo{journal}{Electronic Proceedings in Theoretical Computer Science} \bibinfo{volume}{394}, \bibinfo{pages}{415--431}.
\newblock \DOIprefix\doi{10.4204/eptcs.394.20}.
\bibitem[{Meijer-van~de Griend and Duncan(2023)}]{meijer-vandegriend2020architectureaware}
\bibinfo{author}{Meijer-van~de Griend, A.}, \bibinfo{author}{Duncan, R.}, \bibinfo{year}{2023}.
\newblock \bibinfo{title}{Architecture-aware synthesis of phase polynomials for {{NISQ}} devices}.
\newblock \bibinfo{journal}{Electronic Proceedings in Theoretical Computer Science} \bibinfo{volume}{394}, \bibinfo{pages}{116--140}.
\newblock \DOIprefix\doi{10.4204/eptcs.394.8}.
\bibitem[{Meijer-van~de Griend and Li(2023)}]{meijer2022dynamic}
\bibinfo{author}{Meijer-van~de Griend, A.}, \bibinfo{author}{Li, S.M.}, \bibinfo{year}{2023}.
\newblock \bibinfo{title}{Dynamic qubit routing with {{CNOT}} circuit synthesis for quantum compilation}.
\newblock \bibinfo{journal}{Electronic Proceedings in Theoretical Computer Science} \bibinfo{volume}{394}, \bibinfo{pages}{363--399}.
\newblock \DOIprefix\doi{10.4204/eptcs.394.18}.
\bibitem[{Grover(1996)}]{Grover_1996}
\bibinfo{author}{Grover, L.K.}, \bibinfo{year}{1996}.
\newblock \bibinfo{title}{A fast quantum mechanical algorithm for database search}, in: \bibinfo{booktitle}{Proceedings of the twenty-eighth annual ACM symposium on Theory of computing - STOC ’96}, \bibinfo{publisher}{ACM Press}, \bibinfo{address}{Philadelphia, Pennsylvania, United States}. p. \bibinfo{pages}{212–219}.
\newblock \URLprefix \url{http://portal.acm.org/citation.cfm?doid=237814.237866}, \DOIprefix\doi{10.1145/237814.237866}.
\bibitem[{Kissinger and {\noopsort{griend}}{van de Griend}(2020)}]{kissinger2020cnot}
\bibinfo{author}{Kissinger, A.}, \bibinfo{author}{{\noopsort{griend}}{van de Griend}, A.M.}, \bibinfo{year}{2020}.
\newblock \bibinfo{title}{{{CNOT}} circuit extraction for topologically-constrained quantum memories}.
\newblock \bibinfo{journal}{Quantum Information and Computation} \bibinfo{volume}{20}, \bibinfo{pages}{581--596}.
\newblock \URLprefix \url{http://www.rintonpress.com/journals/doi/QIC20.7-8-4.html}, \DOIprefix\doi{10.26421/QIC20.7-8-4}.
\bibitem[{Li et~al.(2019)Li, Ding and Xie}]{li2019tackling}
\bibinfo{author}{Li, G.}, \bibinfo{author}{Ding, Y.}, \bibinfo{author}{Xie, Y.}, \bibinfo{year}{2019}.
\newblock \bibinfo{title}{Tackling the qubit mapping problem for {{NISQ}}-era quantum devices}, in: \bibinfo{booktitle}{Proceedings of the Twenty-Fourth International Conference on Architectural Support for Programming Languages and Operating Systems}, pp. \bibinfo{pages}{1001--1014}.
\bibitem[{Li et~al.(2022)Li, Wu, Shi, {Javadi-Abhari}, Ding and Xie}]{li2022paulihedral}
\bibinfo{author}{Li, G.}, \bibinfo{author}{Wu, A.}, \bibinfo{author}{Shi, Y.}, \bibinfo{author}{{Javadi-Abhari}, A.}, \bibinfo{author}{Ding, Y.}, \bibinfo{author}{Xie, Y.}, \bibinfo{year}{2022}.
\newblock \bibinfo{title}{Paulihedral: A generalized block-wise compiler optimization framework for {{Quantum}} simulation kernels}, in: \bibinfo{booktitle}{Proceedings of the 27th {{ACM International Conference}} on {{Architectural Support}} for {{Programming Languages}} and {{Operating Systems}}}, \bibinfo{publisher}{{ACM}}, \bibinfo{address}{{Lausanne Switzerland}}. pp. \bibinfo{pages}{554--569}.
\newblock \URLprefix \url{https://dl.acm.org/doi/10.1145/3503222.3507715}, \DOIprefix\doi{10.1145/3503222.3507715}.
\bibitem[{Ljubić(2021)}]{Steiner}
\bibinfo{author}{Ljubić, I.}, \bibinfo{year}{2021}.
\newblock \bibinfo{title}{Solving steiner trees: Recent advances, challenges, and perspectives}.
\newblock \bibinfo{journal}{Networks} \bibinfo{volume}{77}, \bibinfo{pages}{177–204}.
\newblock \DOIprefix\doi{10.1002/net.22005}.
\bibitem[{Lucas(2014)}]{Lucas_2014}
\bibinfo{author}{Lucas, A.}, \bibinfo{year}{2014}.
\newblock \bibinfo{title}{Ising formulations of many {{NP}} problems}.
\newblock \bibinfo{journal}{Frontiers in Physics} \bibinfo{volume}{2}.
\newblock \URLprefix \url{http://journal.frontiersin.org/article/10.3389/fphy.2014.00005/abstract}, \DOIprefix\doi{10.3389/fphy.2014.00005}.
\bibitem[{Naito et~al.(2023)Naito, Hasegawa, Matsuda and Tanaka}]{naito2023isaaq}
\bibinfo{author}{Naito, S.}, \bibinfo{author}{Hasegawa, Y.}, \bibinfo{author}{Matsuda, Y.}, \bibinfo{author}{Tanaka, S.}, \bibinfo{year}{2023}.
\newblock \bibinfo{title}{{{ISAAQ}}: Ising machine assisted quantum compiler}.
\newblock \bibinfo{journal}{arXiv preprint arXiv:2303.02830} .
\bibitem[{Nash et~al.(2020)Nash, Gheorghiu and Mosca}]{nash2020quantum}
\bibinfo{author}{Nash, B.}, \bibinfo{author}{Gheorghiu, V.}, \bibinfo{author}{Mosca, M.}, \bibinfo{year}{2020}.
\newblock \bibinfo{title}{Quantum circuit optimizations for {{NISQ}} architectures}.
\newblock \bibinfo{journal}{Quantum Science and Technology} \bibinfo{volume}{5}, \bibinfo{pages}{025010}.
\bibitem[{Paler et~al.(2021)Paler, Zulehner and Wille}]{paler2021nisq}
\bibinfo{author}{Paler, A.}, \bibinfo{author}{Zulehner, A.}, \bibinfo{author}{Wille, R.}, \bibinfo{year}{2021}.
\newblock \bibinfo{title}{{{NISQ}} circuit compilation is the travelling salesman problem on a torus}.
\newblock \bibinfo{journal}{Quantum Science and Technology} \bibinfo{volume}{6}, \bibinfo{pages}{025016}.
\newblock \URLprefix \url{https://iopscience.iop.org/article/10.1088/2058-9565/abe665}, \DOIprefix\doi{10.1088/2058-9565/abe665}.
\bibitem[{Patel et~al.(2008)Patel, Markov and Hayes}]{patel2008optimal}
\bibinfo{author}{Patel, K.}, \bibinfo{author}{Markov, I.}, \bibinfo{author}{Hayes, J.}, \bibinfo{year}{2008}.
\newblock \bibinfo{title}{Optimal synthesis of linear reversible circuits}.
\newblock \bibinfo{journal}{Quantum Information and Computation} \bibinfo{volume}{8}, \bibinfo{pages}{282--294}.
\newblock \URLprefix \url{http://www.rintonpress.com/journals/doi/QIC8.3-4-4.html}, \DOIprefix\doi{10.26421/QIC8.3-4-4}.
\bibitem[{{Qiskit contributors}(2021)}]{Qiskit}
\bibinfo{author}{{Qiskit contributors}}, \bibinfo{year}{2021}.
\newblock \bibinfo{title}{Qiskit: An open-source framework for quantum computing}.
\newblock \DOIprefix\doi{10.5281/zenodo.2573505}.
\bibitem[{Shor(1994)}]{Shor_1994}
\bibinfo{author}{Shor, P.}, \bibinfo{year}{1994}.
\newblock \bibinfo{title}{Algorithms for quantum computation: discrete logarithms and factoring}, in: \bibinfo{booktitle}{Proceedings 35th Annual Symposium on Foundations of Computer Science}, \bibinfo{publisher}{IEEE Comput. Soc. Press}, \bibinfo{address}{Santa Fe, NM, USA}. p. \bibinfo{pages}{124–134}.
\newblock \URLprefix \url{http://ieeexplore.ieee.org/document/365700/}, \DOIprefix\doi{10.1109/SFCS.1994.365700}.
\bibitem[{Sivarajah et~al.(2021)Sivarajah, Dilkes, Cowtan, Simmons, Edgington and Duncan}]{tket}
\bibinfo{author}{Sivarajah, S.}, \bibinfo{author}{Dilkes, S.}, \bibinfo{author}{Cowtan, A.}, \bibinfo{author}{Simmons, W.}, \bibinfo{author}{Edgington, A.}, \bibinfo{author}{Duncan, R.}, \bibinfo{year}{2021}.
\newblock \bibinfo{title}{t{{|}}ket⟩: a retargetable compiler for {{NISQ}} devices} \bibinfo{volume}{6}, \bibinfo{pages}{014003}.
\newblock \DOIprefix\doi{10.1088/2058-9565/ab8e92}.
\bibitem[{Vandaele et~al.(2022)Vandaele, Martiel and {Goubault de Brugi{\`e}re}}]{vandaele2022phase}
\bibinfo{author}{Vandaele, V.}, \bibinfo{author}{Martiel, S.}, \bibinfo{author}{{Goubault de Brugi{\`e}re}, T.}, \bibinfo{year}{2022}.
\newblock \bibinfo{title}{Phase polynomials synthesis algorithms for {{NISQ}} architectures and beyond}.
\newblock \bibinfo{journal}{Quantum Science and Technology} \bibinfo{volume}{7}, \bibinfo{pages}{045027}.
\newblock \URLprefix \url{https://iopscience.iop.org/article/10.1088/2058-9565/ac5a0e}, \DOIprefix\doi{10.1088/2058-9565/ac5a0e}.
\bibitem[{Winderl et~al.(2023)Winderl, Huang, van~de Griend and Yeung}]{winderl2023architectureaware}
\bibinfo{author}{Winderl, D.}, \bibinfo{author}{Huang, Q.}, \bibinfo{author}{van~de Griend, A.M.}, \bibinfo{author}{Yeung, R.}, \bibinfo{year}{2023}.
\newblock \bibinfo{title}{Architecture-aware synthesis of stabilizer circuits from clifford tableaus}.
\newblock \href{http://arxiv.org/abs/2309.08972}{{\tt arXiv:2309.08972}}.
\bibitem[{Wootters and Zurek(1982)}]{wootters1982single}
\bibinfo{author}{Wootters, W.K.}, \bibinfo{author}{Zurek, W.H.}, \bibinfo{year}{1982}.
\newblock \bibinfo{title}{A single quantum cannot be cloned}.
\newblock \bibinfo{journal}{Nature} \bibinfo{volume}{299}, \bibinfo{pages}{802--803}.
\bibitem[{Wu et~al.(2023)Wu, He, Yang, Shou, Tian, Zhang and Sun}]{wu2019optimization}
\bibinfo{author}{Wu, B.}, \bibinfo{author}{He, X.}, \bibinfo{author}{Yang, S.}, \bibinfo{author}{Shou, L.}, \bibinfo{author}{Tian, G.}, \bibinfo{author}{Zhang, J.}, \bibinfo{author}{Sun, X.}, \bibinfo{year}{2023}.
\newblock \bibinfo{title}{Optimization of {{CNOT}} circuits on limited-connectivity architecture}.
\newblock \bibinfo{journal}{Physical Review Research} \bibinfo{volume}{5}, \bibinfo{pages}{013065}.

\end{thebibliography}

\end{document}